# New determination of the size and bulk density of the binary asteroid 22 Kalliope from observations of mutual eclipses

P. Descamps[1], F. Marchis[1,2,11], J. Pollock[3], J. Berthier[1], F.Vachier[1], M. Birlan[1], M. Kaasalainen[4], A.W. Harris[5], M. H. Wong[2], W.J. Romanishin[6], E.M. Cooper[6], K.A. Kettner[6], P.Wiggins[7], A. Kryszczynska[8], M. Polinska[8], J.-F. Coliac[9], A. Devyatkin[10], I. Verestchagina[10], D. Gorshanov[10]

[1] Institut de Mécanique Céleste et de Calcul des Éphémérides, Observatoire de Paris, UMR8028 CNRS, 77 av. Denfert-Rochereau 75014 Paris, France
[2] University of California at Berkeley, Department of Astronomy, 601 Campbell Hall, Berkeley, CA 94720, USA
[3] Appalachian State University, Department of Physics and Astronomy, 231 CAP Building, Boone, NC 28608, USA
[4] Department of Mathematics and Statistics, Gustaf Hallstromin katu 2b, P.O.Box 68, FIN-00014 University of Helsinki, Finland
[5] DLR Institute of Planetary Research, Rutherfordstrasse 2, 12489 Berlin, Germany
[6] University of Oklahoma, 440 West Brooks, USA
[7] Tooele Utah 84074-9665, USA
[8] Astronomical Observatory, Adam Mickiewicz University, Sloneczna 36, 60-286 Poznan, Poland
[9] Observatoire de la Farigourette 13012 Marseille
[10] Central Astronomical Observatory, Pulkovskoe chaussee 65/1, 196140 St.-Petersburg, Russia
[11] SETI Institute, 515 N. Whisman Road, Mountain View CA 94043, USA

Pages: 59

Tables: 8

Figures: 15

*Corresponding author:*
**Pascal Descamps**
IMCCE, Paris Observatory
77, avenue Denfert-Rochereau
75014 Paris
France

descamps@imcce.fr
Phone: 33 (0)140512268
Fax: 33 (0)146332834

**Abstract**

In 2007, the M-type binary asteroid 22 Kalliope reached one of its annual equinoxes. As a consequence, the orbit plane of its small moon, Linus, was aligned closely to the Sun's line of sight, giving rise to a mutual eclipse season. A dedicated international campaign of photometric observations, based on amateur-professional collaboration, was organized and coordinated by the IMCCE in order to catch several of these events. The set of the compiled observations is released in this work. We developed a relevant model of these events, including a topographic shape model of Kalliope refined in the present work, the orbit solution of Linus as well as the photometric effect of the shadow of one component falling on the other. By fitting this model to the only two full recorded events, we derived a new estimation of the equivalent diameter of Kalliope of 166.2±2.8km, 8% smaller than its IRAS diameter. As to the diameter of Linus, considered as purely spherical, it is estimated to 28±2 km. This substantial "shortening" of Kalliope gives a bulk density of 3.35±0.33g/cm$^3$, significantly higher than past determinations but more consistent with its taxonomic type. Some constraints can be inferred on the composition.

## 1. Introduction

Ten years ago, Pravec and Hahn (1997) suspected that the near-Earth asteroid 1994 AW1 could be a binary system from the analysis of its complex ligthcurve revealed in January 1994. Since that claim, a large number of binary systems have been discovered in all populations of minor bodies in the solar system (Noll, 2006). Extensive and systematic adaptive optics (AO) astrometric follow-up of all known large binary asteroids was performed by our group in order to improve moonlet orbits (Marchis et al., 2003a, 2004). We published a preliminary solution for the orbit of Linus (Marchis et al., 2003b), satellite of the large ($D_{IRAS}$ = 181 km, Tedesco et al., 2002) main belt asteroid 22 Kalliope (Margot et al, 2001). Further AO astrometric observations were carried out up to December 2006, just before the season of mutual events, leading to a significantly corrected orbit solution (Marchis et al., 2008a). Thanks to this updated knowledge of the orbit of Linus, eclipse events were predicted to occur for this binary system from late February through early April 2007 (Descamps et al., 2006). The observation of such events requires a favourable geometry, when the Earth is near the satellite's orbital plane so that eclipses occur at regular intervals. Photometric observation of mutual events is a powerful method to detect and study small asynchronous binaries (Pravec et al. 1998, Mottola and Lahulla, 2000, Ryan et al., 2004, Pravec et al., 2006) as well as doubly synchronous pairs of twin asteroids (Kryszczyńska, 2005, Behrend et al., 2006, Descamps et al., 2007, Marchis et al., 2007a, Kryszczyńska et al., 2008).

Among all asynchronous main belt binary asteroids, the secondary-to-primary size ratio of the 22 Kalliope system is the highest with a value estimated to 0.2 which is considered as the lower photometric detection limit of a binary system in mutual eclipse (Pravec et al., 2006). At the present time, the size of Linus is roughly bound between 20 and 40km, based on measurement of the secondary to primary flux ratio from adaptive optics imaging (Margot and Brown, 2003, Marchis et al., 2003b). With an apparent size of about 25 milliarcseconds (mas), Linus always remains beyond the resolving power of the largest Earth-based telescopes (which is for instance of 55mas for the Keck-10m telescope). With the 2007 favourable circumstances, photometric observations of mutual eclipses enabled us to tightly constrain the physical characteristics of both Linus and Kalliope.

Before tackling the issue of detection of mutual eclipses within the photometric observations collected in 2007 in section 3, we strive in the first place to get an improved polyhedral shape model of Kalliope in section 2, required for a fair modelling of the events. The comprehensive model is fully described in section 4 and fitted to the observations (section 5). The validity of our solution is tested in section 6 with the report of an observation of a stellar occultation of Kalliope and Linus on November 2006. Lastly, we address in sections 7 and 8 some issues as regards the structure and composition of Kalliope and the origin of this binary main-belt asteroid system.

## 2. Improving the topographic shape model of Kalliope

The shape and pole solution are crucial parameters to predict and account for observed photometric rotational lightcurves. Kaasalainen et al. (2002a) derived an initial convex polyhedral shape solution of 22 Kalliope from their noteworthy lightcurve inversion method (Kaasalainen et al., 2001). This model should be considered as a realistic approximation of its highly irregular shape. From November 2006 to March 2007, a long-term photometric follow-up of Kalliope was performed with the 0.4m telescope at Appalachian State University's Rankin Science Observatory, located in western North Carolina.  Images were taken in the R band using an SBIG ST-9e CCD camera and the data were reduced by aperture photometry. The collected lightcurves are displayed on Figure 1. This new set of lightcurves supplied the photometric database of Kalliope in order to update the shape, pole and period solution using the aforementioned inversion method. We took advantage of the edge-on aspect of Kalliope, never observed yet, to improve the current shape model. The derived spin vector solution in J2000 ecliptic coordinates is $\lambda$=195±3° and $\beta$=+7±2° with a sidereal spin period of 4.148199±0.000001hours. The resulting asteroid model is represented as polyhedrons with triangular surface facets. The pear-shaped model for Kalliope is rendered in Figure 2. It turns out to be quite elongated and flattened in a conic-like manner with a theoretical dynamical flattening $J_2$ of 0.19, assuming a uniformly dense body. The scattering of solar light from the surface is synthesized considering the empirical Minnaert's law (1941) with a limb darkening parameter $k$ adjusted to 0.53±0.02, typical of atmosphereless bodies. The surface photometric function is an important determinant of the amount of contrast (darkening) for a given topography. Although the limb-darkening parameter slightly depends on the phase angle, the photometric function of Minnaert provides a reasonable working approximation to more complex, physically motivated models such as the Hapke's function (1981). Nevertheless it becomes somewhat irrelevant at phase angles larger than ~20° (Veverka et al., 1978). Since the present inversion method analysis is mainly based on near-opposition lightcurves and does not take into consideration the modification of the scattering properties with the phase angle, small discrepancies between observed and modelled lightcurves may arise at this range of significant phase angles. In Figure 1, synthetic lightcurves were generated and superimposed on the observations. The evolving peak-to-peak amplitude together with the overall shapes of lightcurves are well accounted for to within about 0.01mag, let alone the lightcurves taken from early March 2007. At this epoch the phase angle was greater than 20° and the departure between our model and the observations reaches 0.02mag at most (effect of such a discrepancy on event modelling is discussed in section 4.1). Furthermore, a posteriori simulations made including the satellite Linus do not show any significant effect, larger than 0.007mag, on the lightcurves (see in Fig. 1 the dash-dotted curve of the March 25 observation). Thereby, neglecting Linus during the lightcurve inversion process does not affect the resulting shape but rather slightly underestimates the limb-darkening parameter by artificially reducing the true amplitude of Kalliope.

## 3. The 2007 mutual events

### 3.1 General description

As the orbit of Linus is nearly located in the equatorial plane of 22 Kalliope, the system undergoes seasons of mutual eclipses and occultations at its equinoxes. This configuration, which occurs every 2.5 years, happened in the northern hemisphere spring of 2007. The observation of such events provides opportunities for very precise astrometry of the satellite and studies of its physical characteristics that are not otherwise possible.

Due to the fast-evolving aspect of the system, as seen by an Earth observer, the season of mutual events lasts only three months. As a result of the distance of 22 Kalliope from the Earth (> 2 AU) and its axial tilt to the ecliptic plane of nearly 90°, mutual eclipses took place in February 2007 and lasted until early April 2007. The proximity of Kalliope to the Sun in May made observations of the mutual occultations difficult.

The brightness of Kalliope ($m_v = 11$) permitted photometric observations with a small aperture telescope. Anomalous attenuation events were predicted to last about 1 – 3 hrs with detectable amplitude ranging from 0.03 to 0.08 magnitudes (Descamps et al., 2006). The magnitude drop during a total eclipse of Linus, the smaller of the two, mainly depends on the relative sizes. An eclipse of Kalliope by Linus is always partial and the decrease in luminosity will also depend on the shape of Kalliope. We estimated that a photometric accuracy of about 0.01-0.02mag was necessary to detect such small photometrical variations.

### 3.2 Observations

The number of observed events depends greatly on the number and geographical distribution of available observers. This is the reason for which an international campaign of observation of these events was set up by the IMCCE team coordinating the efforts to gather observations of as many as possible events[1]. The network of observers is given in the Table 1. Table 2 lists the collected observations which are plotted on Figure 3. Three positive detections have been recorded of which only one was a total eclipse of Linus on March 8th. For the eclipse of Kalliope on March 17th, the ingress was observed at Haute-Provence observatory, France while the egress was recorded in Oklahoma, USA. We merged these two observations into one unique event. Finally, only two complete events were observed. These data are important to assess the durations and thereby constrain the sizes. A third event, an eclipse ingress of Kalliope by Linus, was observed on March 27th, but was not used in the analysis because its incompleteness.

An eclipse can be identified unambiguously if the two components of the binary system have a similar size, but in the case of a small satellite, such as Linus of Kalliope, the effect is so subtle that it may be confused with the primary's lightcurve. To overcome this difficulty, two lightcurves taken a few nights apart are superimposed and subtracted from each other. We will get a positive detection if the residual curve, called the drop curve, reveals a clear light attenuation imputable to a supposed event. The point-to-point subtraction is performed after expanding one of the two lightcurves in Fourier series and interpolating it at the times of the other. Figures 4-6 shows the positive detections. Lightcurves recorded during two closely spaced sessions are overlapped and plotted on the left side of each figure. The corresponding magnitude drop curve, resulting from the subtraction from each other, appears on the right side. Despite the unavoidable presence of noise on the data, prominent light variations are undoubtedly detected and reveal ongoing mutual events within the system of Kalliope.

[1] http://www.imcce.fr/page.php?nav=en/observateur/campagnes_obs/kalliope/index.php

Table 3 summarizes the characteristics of the detected events. For each of them, magnitude drop and times of observed eclipse ingress and egress are reported which may be compared with the expected times, computed from our last orbit solution given in Table 4 (Marchis et al., 2008). The discrepancies are large and may amount to ~2h (180km in position). This is due to the orbit model itself whose the global 1σ RMS error is on the order of 40 mas (70km).The reason lies in a simple preliminary keplerian model used to account for the AO astrometric positions.

### 3.3 **Derivation of the size ratio between Linus and Kalliope from the total eclipse of Linus**

We can take advantage of the full observation of the total eclipse of Linus by Kalliope to estimate the relative sizes of either component of the system. During a total eclipse of Linus, scattered solar light of Linus is not observed on Earth so that the light attenuation is only correlated to the relative size of Linus with respect to Kalliope. The depth, expressed in magnitudes, of the attenuation is related to the ratio of cross-diameters of either component by the simple formula:

$$\frac{F_s}{F_t} = 2.5\log\left(1 + r^2\right) \quad (1)$$

where $r$ is the secondary-to-primary size ratio, $F_s$ the flux of the secondary, and $F_t$ the total flux.

If we apply this formula for an observed attenuation of 0.05mag at the time of disappearance of Linus (Fig.4), assuming the IRAS diameter of Kalliope of 180±4.5km (Tedesco et al., 2002), we get a size ratio of 0.21 corresponding to a size of Linus of 38km. This is in agreement with the size ratio of 0.2, inferred from the flux ratio measurement in the AO observations (Margot and Brown, 2003, Marchis et al., 2003b). Nevertheless, the reliability of such a size determination depends to a great extent on the adopted size of Kalliope. The mutual events observations provide a unique opportunity to get an independent measurement of the very size of Kalliope which will mainly decide on the duration of an event. Furthermore, if we look at the drop curves, we must pay attention to the fact that they do present neither the same pattern nor similar amplitudes. They are not at all reminiscent of the reversed bell curves observed in the case of classical mutual events within planetary satellite systems. Quite obviously, the shape of Kalliope plays a key role in the way the light attenuation is taking place. Therefore, we should first address the modelling of an eclipse phenomenon, involving a non-spherical primary body, in order to derive a trustworthy new assessment of the size of Kalliope.

## 4. Modelling the mutual eclipses

### 4.1 Description of the synthetic model

From the study presented in section 2, we have a precise knowledge of the rotation and shape of the primary. The shape of the secondary cannot be constrained by mutual eclipse observations owing to its smallness relatively to the primary and will be consequently considered as purely spherical. Another point of importance lies on the precise localization of the shadowed area on the surface of Kalliope whenever it is eclipsed by Linus. This is achieved if we conveniently consider that an eclipse is an occultation from the standpoint of a Sun observer. From the Sun, the cross sections of the "occulting" and "occulted" bodies draw outlines. Facets of the "occulted" body (the eclipsed body for an Earth observer) which are located inside or on the border of the occulting body outline do not receive any solar light and their scattered flux is equal to zero. After going back to the initial frame facing the observer – which is that of the tangent plane of the observation – we have to retrieve the eclipsed facets and to carry out a summation of the scattered flux over all visible facets including the eclipsed facets. In our model the penumbral annulus is not taken into account because each body is located very close from each other so that the penumbral width is negligible. Light travel time between each body is likewise neglected. The model provides the values of the quantities $F_1^e$, $F_2^e$, $F_1$ and $F_2$ which stand for the fluxes of Kalliope (subscript 1) and Linus (subscript 2) in and out eclipse. The magnitude drop of a partial eclipse of Kalliope by Linus (2E1) or of an eclipse of Linus by Kalliope (1E2) is straightforwardly provided by the following formula:

$$\Delta F_{1E2} = \frac{F_1 + F_2^e}{F_1 + F_2} \quad (2)$$

$$\Delta F_{2E1} = \frac{F_1^e + F_2}{F_1 + F_2} \quad (3)$$

The eclipse of Linus is said *total* if $F_2^e=0$. Figure 7 shows the apparent configuration of the system generated at three times of the partial eclipse of Kalliope by Linus on March 17th. The dark area on Kalliope, which is the shadow of Linus, covers an irregular region, slightly tilted over the cylindrical shadow. Thus the umbra obliquely falls upon the surface, causing, for an Earth-based observer, a distortion of the shadow which tends to encompass a much larger area than we would have on a purely spherical Kalliope. On the other hand, the irregular shape of Kalliope makes that its cross-section strongly departs from that corresponding to its effective diameter. Accordingly, the part of flux which is removed from the total collected light is enhanced which explains why we observe decrease in magnitude as large as 0.08mag when the shadow of Linus is falling on Kalliope.

We saw in section 2 that our updated shape model of Kalliope was able to reproduce observed lightcurve to within about 0.02mag. A question which may then come to mind is whether this topographic model is capable of simulating attenuations as faint as 0.05-0.08 magnitudes. Let $\delta F_1$, much less than $F_1$, be the error in the Kalliope flux estimation due to its approximated shape. Take now $F_1+\delta F_1$ and $F_1^e+\delta F_1$, and substitute for $F_1$ and $F_1^e$ into Eq. [3] (same reasoning with the Eq. [2]). We have

$$\Delta \tilde{F}_{2E1} = \frac{F_1^e + \delta F_1 + F_2}{F_1 + \delta F_1 + F_2} = \Delta F_{2E1}\left(1 + \delta F_1\left(\frac{F_1 - F_1^e}{(F_1 + F_2)(F_1^e + F_2)}\right) + O\left(\frac{\delta F_1^2}{F_1^2}\right)\right) \quad (4)$$

Hence, assume that $F_1^e \simeq F_1 - \delta F_1$ and $F_2 \leq F_1$

$$\Delta \tilde{F}_{2E1} \approx \Delta F_{2E1}\left(1 + \frac{\delta F_1^2}{F_1^2} + O\left(\frac{\delta F_1^2}{F_1^2}\right)\right) \quad (5)$$

Taking into account that the error expressed in magnitude is given by $2.5\log(1+\delta F_1/F_1)\sim0.02$mag, the error in the magnitude drop then amounts to $\sim2.5\log(1+\delta F_1^2/F_1^2)\sim0.0004$mag, which proves the ability of our shape model to account for the drop curve during an event. As a rule, if we want to have a fair representation of the magnitude drop, it is sufficient to have a shape approximation not yielding photometric departure from the observed lightcurves, no matter what the aspect and phase angles, by more than ~0.05mag (yielding 0.002mag on the drop curve). This condition rules out a trivial triaxial ellipsoid as a valid shape model to simulate the light loss during an event involving an irregular body.

### 4.2 Photometric effect of the eclipse parameters

This part is aimed at scrutinizing the sensitivity of the drop curve against the main parameters involved in an eclipse event. The free parameters taken into account are the size of bodies, the orbital pole, the shape model of Kalliope and the Minnaert limb-darkening parameter. Despite their unavoidable correlations, this study is undertaken to highlight the characteristic dominant effect of each parameter. It is carried out with the event on March 8$^{th}$. Apart from our refined topographic shape model, we considered another shape model provided by its current ellipsoidal figure (a/b=1.33, b/c=1.27, de Angelis, 1995). The effects on the duration, the amplitude and the shape of the drop curve of the event are visible on Figure 8.

Fig. 8F shows that the most conspicuous effect arises from the shape model of Kalliope. For a same orientation and effective diameter, the ellipsoidal model cannot account not only for the irregular profile of the drop curve but also for the amplitude which is then significantly underestimated. In other words, using an ellipsoidal model will yield an overestimation of the Linus size, required to compensate the lack of amplitude. As to the influence of the Minnaert limb-darkening parameter (Fig. 8E), it is negligible given our estimated value of 0.53±0.02 (see section 2).

Based on these results, we will proceed in the rest of this study by definitely adopting the polyhedral shape solution of Kalliope obtained in section 2. As far as the size effects are concerned, they are consistent with what it is expected in terms of amplitude and duration variations, namely, the smaller the Kalliope size, the shorter the duration and the weaker the amplitude (Fig. 8A). However, the Linus size has no measurable effect on the duration, but only on its amplitude (Fig. 8C & 8D). Basically, the most significant changes in the shape of the drop curve come from the orbital pole orientation. The influence of each ecliptic coordinate has been explored separately and the prominent effect is decidedly on the very form of the drop curve which may markedly evolve with only slight variations on either coordinate.

Lastly, from this series of trials, we can now draw some general rules to iteratively process each observed event and achieve a rough starting solution for each parameter:

1. Determining the best Linus orbital plane orientation able to accurately match the general form of the drop curve.
2. Fitting the size of Kalliope to the duration.
3. Fitting the size of Linus to the amplitude.

## 5. General solution

### 5.1 Physical solution for Kalliope and Linus

Practically, we perform a full grid analysis over the multidimensional space of the free parameters (ecliptic coordinates of the orbit pole, equivalent radii of Kalliope and Linus) using a goodness-of-fit criterion taken as the averaged differences between simulations and observations. The criterion Θ is defined as:

$$\Theta(mag) = \sqrt{\frac{\sum_{i=1}^{n}\left(O_i - C_i\right)^2}{n}} \qquad (6)$$

Where n is the data number, and $O_i$ and $C_i$ are observed and calculated magnitudes.

Before each fit, the spin axis of Kalliope was finely determined from the corresponding reference lightcurve. We found the following J2000 ecliptic coordinates, (192±2°, +7±1°) on March 6th and (196±2°, +6±1°) on March 18th. The grid search in the parameter space was performed inside a region surrounding an initial guess of the best solution provided by the three rules described in the previous section. The best solution is reached for the event on March 8th by minimizing the goodness-of-fit criterion with Θ=0.008mag. Then we derive a precise estimate of the uncertainty on the Kalliope effective diameter by investigating the relevance of the event duration to the size of Kalliope (Fig. 9). Given an accuracy of 1 minute of time on the ingress and egress times determination (Table 3), i.e. 2 minutes in the duration assessment, we derive an effective diameter of Kalliope of 166.2±2.8km which turns out to be approximated by a triaxial ellipsoid with semi-major axes a=117.5km, b=82km, c=62km. The Linus diameter is then fitted to 28±2km. With this new value of the effective diameter of Kalliope, we derive a visible albedo $p_v$ =0.17[2], higher than its previous value of 0.12, for an absolute magnitude H=6.45. This albedo is in agreement with the mean albedo of M-type asteroids (Belskaya and Lagerkvist, 1996). In order to ascertain the reliability of these results, we successfully fitted the event on March 17th with the same size parameters and Θ=0.014mag, giving a high degree of coherence and confidence to our size derivation. The best-fitted drop curves are displayed on Figure 10. As for the Linus orbital plane orientation, their J2000 ecliptic coordinates are fitted to (194±1°, -4±1°) on March 8th and (198±1°, -5±1°) on March 17th. Although it implies a moderate inclination over the spin axis of Kalliope of ~11°, yielding an expected nodal precession rate of 0.15°/day. The physical interpretation of such a motion in the spin and orbital axes should be cautiously considered and deserves further observational confirmations and theoretical investigations.

[2] If the size D of an asteroid is known and its absolute magnitude H, its geometric visible albedo $p_v$ can be derived from $D(km) = \frac{1329}{\sqrt{p_v}}.10^{-H/5}$

It is worth noting that we get a new secondary-to-primary size ratio of 0.169±0.006, instead of 0.21 derived from AO observations. This lower size ratio is likely due to the fact that Kalliope is always resolved in AO images and consequently its flux is spread over its apparent size, differently than it would be with a point-like stellar profile. A consequence of that photometric bias is the overestimation of the secondary size if we simply adopt the previously published IRAS diameter for the primary. Besides, in the case of an irregular body such as Kalliope for instance, the expected theoretical magnitude drop, given by Eq. [1], is basically inadequate to give a fair idea of the real level of the light attenuation. With the new size ratio, the decrease in magnitude assuming spherical primary would be but 0.03mag, lower than what was really observed by a factor of ~3. This difference underlines the enhancement of light loss due to the degree of non-sphericity of the primary. As a consequence, binary systems with much smaller size ratio than 0.2 may be studied through mutual events observations. Adopting a detection threshold of ~0.02mag, we may infer a limiting secondary-to-primary size ratio of ~0.07, necessary to detect mutual events within a binary asteroidal system.

### 5.2 Astrometry of the events

It is straightforward from our photometric model to infer ancillary astrometric relative positions of Linus in the tangent plane of the observation at the times of eclipse ingress and egress. The astrometric positions labelled in Table 5 are extremely accurate with a typical error of 1 mas and 4 mas in X and Y axes respectively. A significantly more advanced dynamical model, taking into account the non-spherical gravitational field of the primary as well as the moderate inclination of the orbital plane over the equator of Kalliope, is necessary to accurately account for the motion of Linus.

## 5.3 Comparison with IRAS thermal observations

Tedesco et al. (2002) reported an estimated diameter for 22 Kalliope of 181+/-5 km, based on radiometric measurements made during four observations by the IRAS telescope; this value is surprisingly ~8% larger than our calculated diameter. On the basis of this larger average diameter result, Marchis et al. (2003b) reported a very high porosity (~60%) for the primary, which is difficult to explain with any realistic model of the asteroid interior.

We retrieved and reanalyzed individually all the relevant IRAS data using the NEATM algorithm described by Harris (1998). We estimated the size, albedo and η for each observation separately (see Table 7). In the NEATM, the parameter η is a modeling parameter, often referred to as the beaming parameter, which is adjusted together with the diameter to obtain the best fit to the observed thermal continuum. By incorporating a variable η value, the NEATM takes into account the effects of thermal inertia and surface roughness (unlike the standard thermal model on which the original IRAS analysis was based). We discarded IRAS observation #2 because it gives an abnormal η value, possibly due to a background star contamination in the 12 μm measured flux. The average NEATM diameter (177+/-4 km with $p_v$=0.15 and η=0.753+/-0.042) is, however, still significantly larger than the one we report in this paper.

Figure 11 shows the geometry of Kalliope's primary at the times of the IRAS data (June 15 and 19, 1983), calculated using the pole solution and 3D-shape described in this work. The asteroid is seen very close to pole-on, meaning that it was observed while showing its largest apparent surface. Using our model mean diameter of 166.2 km, the calculated effective diameter at the time of the IRAS sightings (~187 km) is very close to the average diameter derived from the IRAS data using NEATM (the effective diameter is the diameter of a sphere with equivalent projected area to that of the asteroid). If we also include a possible thermal contribution of Linus (R=14 km), the calculated effective diameter of the primary is reduced to ~180 km is now in agreement with the NEATM diameter.

In conclusion, it is evident that the diameter of Kalliope was overestimated based on the IRAS observations because the few IRAS radiometric measurements were made while Kalliope was viewed pole on. Our measured effective diameter of 166.2 km is in agreement with the NEATM diameter, taking into account the pole solution and 3D-shape reported in this work.

## 6. First stellar occultation by the satellite of an asteroid

On November 7, 2006, Japanese observers recorded the first predicted occultation of a star, TYC2 1886012061, by the satellite of an asteroid, Linus (Soma et al., 2006). The event was announced by Berthier et al. (2006) on the basis of our orbital model of Linus (Marchis et al., 2008). The geodetic coordinates and altitudes of the observing sites as well as the names of the observers are listed in Table 7. This observation not only confirmed the reliability of our orbit solution but also provided a direct measurement of the sizes of Linus and Kalliope.

From a sufficient number of immersion and emersion timings of a stellar occultation, the radius and ellipticity of the occulting body can be accurately determined. In this method, the coordinates of the observed occultation points are projected onto a common plane that passes through the center of the Earth and lies perpendicular to the line between the centers of the star and the occulting body. After the occultation points are projected onto the occultation plane, we obtain several chords to which an elliptical figure may be fitted. Each chord is a single line across the shadow cast by the asteroid or its satellite as it moves over the observer. Adopting INPOP06 (Fienga et al., 2006) as the reference planetary theory, the ephemeris of Kalliope provides a geocentric apparent velocity of 17.46 km/s at the epoch of the occultation with an accuracy better than one meter per second. Immersion and emersion timings are reported in Table 8 and inferred chords of both bodies are plotted as solid lines on Figure12. The dashed lines show the negative observations (i.e. no disappearance of the star reported by the observer). After adjusting the irregular outline of Kalliope to the observed chords, we may assess an astrometric shift of 80 ±10 km between the projected center of Kalliope and its predicted position materialized by the origin of the frame. Note that, at the epoch of the occultation, the current ephemeris uncertainty (1-sigma RMS) of Kalliope provided by ASTORB (Bowell, 2007) is 30 mas or 39 km in the occultation plane and the total uncertainty on the position of the star given by the TYCHO-2 catalogue is 49 mas or 57 km in the occultation plane. With regard to Linus, the astrometric residual is 58 ±12 km (45 ±9 mas) at the mean time of the occultation (2006-11-07T19:49:26 UTC).

Despite the limited coverage of the limb profile by the recorded chords (Fig. 13), an apparent elliptical figure of Linus has been fitted with semimajor axes of 16±5 km and 14 ±1 km, giving an equivalent radius of 15 ±3 km, in good agreement with our previous determination. The quoted errors reflect only the insufficient number of chords and do not allow to accurately constraining the global shape.

The small number and the sparse distribution of the chords along the cross-section of Kalliope provide a limited sketch of its limb profile. Nevertheless, from our topographic shape model of Kalliope (see section 2), we can homothetically expand or shrink its projected profile onto the occultation plane along the line of sight, in order to adjust its best equivalent radius. In Figure 14, two limb profiles, corresponding to the Kalliope equivalent radius of 83.1km, derived in the present work, and to its IRAS radius (90.5 km), have been superimposed to the observed chords. The adjustment is performed from the lower points, which correspond to the emersion of the star, the timings of which are usually more accurate and less affected by the reaction time of an observer. The negative observations, located at the bottom right, put alike strong constraints on positioning the profiles. The Kalliope shape-derived profile with an equivalent radius of 83.1 km better encompasses the observed chords with a mean residual between observed and computed chord lengths of 4.6 ±7.3 km versus 8.1 ±11.0 km as derived from the 90.5 km IRAS radius. However, a significant deviation of the order of 20 km (or 1.1 second of time) from the projected profile is apparent at the immersion timings of chords 5 and 7. This departure cannot be reasonably explained considering only a timing error from the observer. It rather indicates the possible existence of a prominent non-convex topographic feature on Kalliope or a slightly deformed shape near its narrowest tip. Dŭrech and Kaasalainen (2003) mentioned that nonconvexities of an asteroid can be recovered from lightcurve inversion only from observations at sufficiently high solar phase angles ($\alpha$>60°) when shadowing effects from nonconvexities become important. From the Figure 2 we may notice that the main feature of the overall shape is a cone-like structure.  As already noted by Kaasalainen et al. (2002b) for the asteroid 44 Nysa, the sides of the cone are alike probably concave because long, straight stretches on the convex solution, as it is the case for Kalliope, usually indicate nonconvexity.  Further observation of such similar stellar occultations should confirm and constrain this hypothesis.

## 7. Revised physical characteristics of Kalliope

From the refined orbit solution of Linus in its motion around Kalliope, the bulk density of Kalliope can be inferred. From the Keplerian mean motion $n_0$ given in Table 4 we can derive the total mass $M_T$ of the system according to the Kepler's third law:

$$n_0^2 a_0^3 = GM_T = GM_K(1+q) \quad (7)$$

where $G$ is the gravitational constant, $M_T$ the total mass, $M_K$ the mass of Kalliope, $a_0$=1099±11km the semi-major axis and q=0.0047, the mass ratio between Linus and Kalliope assuming a same bulk density for both of them. We obtain a mass for Kalliope of $8.16 \pm 0.26 \times 10^{18}$ kg. Given our diameter of Kalliope of 166.2±2.8 km we calculate a bulk density $\rho = 3.35 \pm 0.33 g/cm^3$ which is nearly twice the density derived from previous determinations based on the IRAS diameter (Marchis et al., 2003b, Margot and Brown, 2003). It is worth emphasizing that our semi-major axis differs significantly from the one of Margot and Brown (2003). Such a discrepancy has obviously consequences on mass and density determination. Basically, it chiefly depends on the number of astrometric data to which the orbit solution is fitted. Our solution was fitted to 41 astrometric positions ranging from August 2001 to December 2006 (Marchis et al., 2008) while that of Margot and Brown (2003) utilized 9 observations spanning a little more than one year between August 2001 and December 2002. Furthermore, their orbit pole solution was estimated to be (196.2, +3.2) in J2000 ecliptic coordinates, departing by ~7° from our solution (Table 4) which was confirmed in the present work.

In a recent work, Descamps and Marchis (2008) set forth the idea that most of the main belt binary systems with a large primary, which can be considered as rubble-piles, might originate in a rotational fission process chiefly governed by the level of the internal friction. Adopting their notations, we update the normalized spin rate of Kalliope to Ω=0.425±0.005 and the non-sphericity parameter λ=1.494, which is the ratio between the moment of inertia of the body with respect to its spin axis and the moment of inertia of the equivalent sphere. The specific angular momentum of the system is then be computed from a handy formula given in the appendix A of their paper, H=0.269±0.008[3], akin to the average value of the observed main belt binary systems. In other words, despite the new physical properties of Kalliope derived in this work, the main conclusions as regards to the total angular momentum hold true. The binary system of Kalliope could likely be the outcome of rotational fission provided that its internal structure is of rubble pile. This assumption would require that the macroscopic porosity to be on the order of ~30%, corresponding to the adopted lower porosity for loose rubble or soils (Britt et al. 2002). The assignment of Kalliope to the “M” type in the Tholen taxonomy (Tholen, 1989) or Xe type in the SMASS II taxonomy (Bus and Binzel, 2002) is based on moderate albedo (~0.1-0.3), the lack of identified mineral absorption features in the visible and near-infrared reflectance spectra and a flat or slightly red spectral curve across 0.3-2.5μm wavelength range. Interpretation of such spectra is very challenging because of the lack of diagnostic features. The apparently featureless spectra of M-asteroids led to suggest that either enstatite chondrites or NiFe meteorites were the best analogs for the M-type asteroids (Gaffey and McCord, 1979). Recently, Rivkin et al. (1995, 2000) suggested a subgroup of M-class asteroids (called the W class) containing those with an absorption feature at 3μm diagnostic of water of hydration although its reliability is questioned (Gaffey et al., 2002). In other terms, members of this new class are

---

[3] $$H=\frac{q}{(1+q)^{\frac{13}{6}}}\sqrt{\frac{a(1-e^2)}{R_p}}+\frac{2}{5}\frac{\lambda_p}{(1+q)^{\frac{5}{3}}}\Omega+\frac{2}{5}\lambda_s\frac{q^{\frac{5}{3}}}{(1+q)^{\frac{7}{6}}}\left(\frac{R_p}{a}\right)^{\frac{3}{2}}$$

Where $q$ is the secondary-to-primay mass ratio, $a$ the semimajor axis, $R_p$ the primary radius and $e$ the eccentricity. The normalized spin rate is defined by $\Omega = \omega^2/\pi\rho G$ (ω is the orbital angular velocity; ρ is the density assumed to be equal for the two components; G is the gravitational constant).

hydrated asteroids with surface composition inconsistent with pure metallic iron. Kalliope has been identified as one of them. These authors concluded that the W class asteroids likely have a mineralogy akin to enstatite chondrites ($3.36 g/cm^3$) or carbonaceous chondrite-like ($3.05$-$3.75 g/cm^3$) with a relatively large fraction of high-albedo hydrated salts at the surface. Given our revised bulk density of Kalliope, the candidate meteoritic analogs should have a grain density no less than ~$3.4 g/cm^3$. Thus adopting an enstatite composition would lead to make Kalliope a giant monolithic boulder with an almost zero porosity. This seems not realistic regarding its specific angular momentum, previously derived, putting it among other main-belt asteroids hitherto considered as rubble-piles. The only relevant analogs may therefore be more likely either iron meteorites with almost zero microporosity (grain density of $6.99$-$7.59 g/cm^3$) or stony-iron meteorites with silicate-bearing iron ($5.0 g/cm^3$) (Prinz et al., 1982, Rivkin et al., 1995, 1997, Magri et al., 2001, Hardersen et al., 2005, Birlan et al., 2007). The degree of silicate inclusions varies significantly but can compose up to ~50% by volume (Prinz et al., 1984). Presence of metal suppresses silicate absorption features very significantly (Feierberg et al., 1982). Cloutis et al. (1990) showed that at least 10% of orthopyroxene or 25% of olivine must be present to be spectrally resolvable. The lack of diagnostic feature in the ~0.9µm spectral region (Marchis et al., 2008b) rules out stony-irons such as mesosiderites ($4.16$-$4.22 g/cm^3$) or pallasites ($4.82$-$4.97 g/cm^3$). As a rule, present state of knowledge on the composition of Kalliope does not allow to firmly assess the macroscopic porosity from the surface grain density. Wilson et al. (1999) showed that post-impact, gravitationnally reaccreted asteroids should have porosities of ~20-40%. Accordingly, acknowledging a rubble-pile structure of Kalliope from its specific angular momentum, such a range of porosities implies a grain density in the $4.2$-$5.8 g/cm^3$ range, suggestive of a mixture of NiFe metal-rich with inclusion of silicates.

## 8. **Discussion of origin**

In Figure 15 we have plotted tidal evolution time scales as a function of the relative separation, and mass ratio (Weidenschilling et al., 1989), we may put corresponding values derived for Kalliope, $a/R_p=13.16$ and $q=0.0047$. We have adopted $\mu Q \approx 3x10^{13}$dynes/cm$^2$. The specific energy dissipation function Q is generally ~100. As to the coefficient of rigidity $\mu$, it ranges from $3x10^{10}$dynes/cm$^2$ (ice) to $5x10^{11}$dynes/cm$^2$ for uncompressed but well-consolidated rocky materials (For instance granite has $\mu \sim 3x10^{11}$dynes/cm$^2$). Moderately fractured carbonaceous asteroids (such as Phobos) have $\mu \sim 10^{10}$dynes/cm$^2$. If we adopt $\mu$ to be at least $10^{11}$dynes/cm$^2$ for a stony-iron Kalliope, we can bind the evolution time, assuming initial $a/R_p=1$, between ~1.0 and ~3.0 billion years, making Kalliope as likely one of the most primitive main belt binary systems. This rough estimation can be connected with the supposed, although controversial, cataclysmic bombardment which occurred at around 3.9Gyr in the inner solar system. A recent model tentatively explains this so called, late heavy bombardment as the result of the readjustment of the orbits of Giant Planets in the outer solar system (Tsiganis et al., 2005, Gomes et al., 2005). The process would have led to a massive delivery of planetesimals to the inner Solar system, and consequently the disruption of the asteroid belt. A loosely consolidated proto-Kalliope may have experienced off-center impacts able to deliver enough angular momentum in order to reach the threshold of instability for a rubble-pile asteroid with a given friction angle (Holsapple, 2004). According to our derived value for the global specific angular momentum of 0.27, this threshold would be reached for a friction angle in the 11-14° range (Descamps and Marchis, 2008). Weidenschilling et al. (1989) already invoked such a mechanism to account for the formation of binary asteroids by rotational fission but the required impact energy would exceed the binding energy of an asteroid so that the probability an asteroid would survive is near zero. Actually, this catastrophic fate is no longer true for porous asteroids. Recent laboratory experiments and scaling arguments have shown that because a fractured, highly porous asteroid transmits impact energy so poorly, it is much more difficult to break up than a monolith (Benz and Asphaug, 1999, Housen et al., 1999, Housen and Holsapple, 2003). A new cratering mechanism occurs producing the compaction of the target material with ejection velocities substantially lower. In large impacts, most of the ejected material never escapes the crater. Therefore, a porous proto-Kalliope could survive a tremendous impact which would impart enough angular momentum. This process will bring the proto-kalliope to an unstable rotational state able to give rise to a binary system assuming a Mohr-Coulomb model for the internal structure of a rubble-pile (Holsapple, 2004). The supposed nonconvexity on Kalliope we alluded to from the stellar occultation observation (section 6 and Fig.14) could be the result of such a tremendous impact. This interesting scenario remains extrapolation based on these data. We do believe that close-up studies of Kalliope and Linus surfaces that could be provided by a dedicated space mission will certainly help us to understand how Kalliope binary system formed.

## 8. Conclusion

This work provides new insights about the physical characteristics of Kalliope. The new size measurement of Kalliope does not support its status as the one of the most porous objects observed so far. These new conclusions have been achieved after combining the results of various observational techniques. Firstly, adaptive optics observations (Marchis et al., 2003b, Marchis et al., 2007), are the cornerstone of predictions of mutual events and stellar occultations (accessible to amateur astronomers), as they provide precise knowledge of the orbit of the secondary. Secondly, continuous lightcurve observations, combined with a robust and powerful inversion method (Kaasalainen et al., 2001), enable the construction of a robust 3-dimensional shape of the primary. These different sets of observational data have allowed an accurate and self-consistent model of the Kalliope system to be developed, with important implications for its physical characteristics. We can thus provide a global and accurate solution for Kalliope in terms of size, shape, spin axis, and scattering properties. The main outcome is a revised effective diameter for Kalliope of 166.2±2.8km with a secondary, Linus, 28±2km in diameter. These results are confirmed by a new analysis of IRAS existing radiometric data and the recent observation of a stellar occultation by the system of Kalliope. As a consequence, we derive a bulk density of $3.35 \pm 0.33 g/cm^3$, much higher that derived previously.

## Acknowledgements

This work was supported equally between the National Science Foundation Science and Technology Center for Adaptive Optics, managed by the University of California at Santa Cruz under cooperative agreement No. AST-9876783 and the National Aeronautics and Space Administration issue through the Science Mission Directorate Research and Analysis Programs number NNX07AP70G. We are thankful to our anonymous reviewers for the quality of their comments and for their willingness to clarify the manuscript.

# Tables

## Table 1:

List of the observers, theirs facilities as well as the observational conditions.

| Observers | Observatory | Aperture (m) | Exposure Time and filter (s) | Average seeing (arcsec.) |
|---|---|---|---|---|
| Coliac, J.-F. | Farigourette Observatory 5°27'21"E 43°18'38"N | 0.20 | 10 (Clear) | 3 |
| Descamps, P., Vachier, F. | Haute-Provence Observatory IAU code #911 | 1.20 | 30 (R) | 3 |
| Deviatkin, A., Verestchagina, I.A., Gorshanov, D.L. | Pulkovo Observatory IAU code #084 | 0.32 | 120 (B)<br>60 (V)<br>30 (R)<br>60 (I) | 5 |
| Kryszczynska, A., Polinska, M. | Borowiec IAU code #187 | 0.40 | 180 (Clear) | 5 |
| Marchis F., Wong, M.H. | Lick Observatory IAU code #662 | 1.00 | 10(R) | 1.4 |
| Pollock, J. | Appalachian State University, Rankin Science Observatory 81°40' 54"W 36°12' 50"N | 0.40 | 30(R) | 3 |
| Romanishin, W., | University of Oklahoma Observatory IAU code #30 | 0.40 | 60 | 3.5 |
| Wiggins P. | IAU code #718 | 0.35 | 15 (Clear) | 2 - 4 |

**Table 2:** Summary of the observations. The object 1 is referred to as Kalliope and the object 2 as Linus. Each event is identified by its Id number. The columns “Exp.” and “Obs.” give respectively the expected event inside the observational window and the observed event. “2E1” stands for an eclipse of Kalliope by Linus. Some predicted events may turn out not to be observed due to the relative inaccuracy of the actual orbit solution of Linus on the order of 220km (or ~2h30 in time shift) at worst..

| Date | Id | UT Start | UT End | Duration | Exp. | Obs. | Observer |
|---|---|---|---|---|---|---|---|
| 11/26/06 | 1 | 03:14:06 | 09:53:11 | 06:39:05 | | - | J. Pollock |
| 11/27/06 | 2 | 23:58:50 | 06:08:20 | 06:09:30 | | - | J. Pollock |
| 12/02/06 | 3 | 01:04:16 | 11:49:14 | 10:44:58 | | - | J. Pollock |
| 12/13/06 | 4 | 23:45:05 | 11:23:18 | 11:38:13 | | - | J. Pollock |
| 12/15/06 | 5 | 00:45:35 | 11:16:15 | 10:30:40 | | - | J. Pollock |
| 12/19/06 | 6 | 00:40:36 | 07:59:13 | 07:18:37 | | - | J. Pollock |
| 01/27/07 | 7 | 00:02:00 | 07:54:11 | 07:52:11 | | - | J. Pollock |
| 02/22/07 | 8 | 02:08:32 | 05:31:55 | 03:23:23 | | - | W. Romanishin et al. |
| 02/27/07 | 9 | 03:42:18 | 05:43:46 | 02:01:28 | **2E1** | - | W. Romanishin et al. |
| 02/28/07 | 10 | 02:31:55 | 07:17:57 | 04:46:02 | | - | P. Wiggins |
| 03/03/07 | 11 | 02:32:37 | 06:46:43 | 04:14:06 | | - | P. Wiggins |
| 03/04/07 | 12 | 03:16:28 | 07:36:09 | 04:19:41 | | - | P. Wiggins |
| 03/05/07 | 13 | 02:53:46 | 07:39:00 | 04:45:14 | | - | P. Wiggins |
| 03/06/07 | 14 | 00:09:23 | 05:08:12 | 04:58:49 | | - | J. Pollock |
| 03/07/07 | 15 | 02:17:56 | 05:31:54 | 03:13:58 | | | W. Romanishin et al. |
| 03/07/07 | 16 | 23:58:50 | 06:08:20 | 06:09:30 | **1E2** | **1E2** | J. Pollock |
| 03/08/07 | 17 | 01:55:50 | 05:30:34 | 03:34:44 | **1E2** | **1E2** | W. Romanishin et al. |
| 03/07/07 | 18 | 02:56:17 | 08:39:04 | 05:42:47 | | - | P. Wiggins |
| 03/09/07 | 19 | 02:23:30 | 08:28:05 | 06:04:35 | | - | P. Wiggins |
| 03/10/07 | 20 | 00:00:20 | 04:00:04 | 03:59:44 | | - | J. Pollock |
| 03/10/07 | 21 | 02:08:34 | 05:33:19 | 03:24:45 | | - | W. Romanishin et al. |
| 03/11/07 | 22 | 03:14:22 | 08:00:27 | 04:46:05 | | - | P. Wiggins |
| 03/12/07 | 23 | 06:05:33 | 07:35:49 | 01:30:16 | | - | P. Wiggins |
| 03/16/07 | 24 | 18:34:31 | 00:58:15 | 06:23:44 | | **2E1** | P. Descamps et al. |
| 03/17/07 | 25 | 01:35:36 | 05:14:20 | 03:38:44 | **2E1** | **2E1** | W. Romanishin et al. |
| 03/17/07 | 26 | 04:26:28 | 07:22:53 | 02:56:25 | **2E1** | **-** | F. Marchis et al. |
| 03/18/07 | 27 | 01:36:28 | 05:13:45 | 03:37:17 | | | W. Romanishin et al. |
| 03/18/07 | 28 | 03:03:07 | 06:03:56 | 03:00:49 | | | F. Marchis et al. |
| 03/18/07 | 29 | 02:32:43 | 07:11:53 | 04:39:10 | | - | P. Wiggins |
| 03/19/07 | 30 | 02:51:20 | 07:27:04 | 04:35:44 | | - | P. Wiggins |
| 03/23/07 | 31 | 02:34:44 | 07:40:41 | 05:05:57 | | - | P. Wiggins |
| 03/25/07 | 32 | 00:25:40 | 05:01:01 | 04:35:21 | | - | J. Pollock |
| 03/25/07 | 33 | 02:22:39 | 06:54:44 | 04:32:05 | | - | P. Wiggins |
| 03/27/07 | 34 | 19:17:02 | 22:50:18 | 03:33:16 | | **2E1** | J.F. Colliac |
| 03/27/07 | 35 | 20:43:36 | 22:58:09 | 02:14:33 | **2E1** | | A. Deviatkin et al. |
| 03/27/07 | 36 | 20:46:01 | 23:00:00 | 02:13:59 | **2E1** | | A. Deviatkin et al. |
| 03/27/07 | 37 | 20:47:34 | 23:01:09 | 02:13:35 | **2E1** | | A. Deviatkin et al. |
| 03/27/07 | 38 | 20:48:49 | 23:02:45 | 02:13:56 | **2E1** | | A. Deviatkin et al. |
| 03/27/07 | 39 | 19:08:35 | 23:56:14 | 04:47:39 | | **2E1** | A. Kryszczynska et al. |
| 04/03/07 | 40 | 03:03:20 | 06:47:46 | 03:44:26 | | - | P. Wiggins |
| 04/04/07 | 41 | 02:40:22 | 04:52:29 | 02:12:07 | | **-** | P. Wiggins |
| 04/06/07 | 41 | 02:40:06 | 06:37:52 | 03:57:46 | | - | P. Wiggins |
| 04/07/07 | 42 | 02:44:48 | 06:29:21 | 03:44:33 | | - | P. Wiggins |
| 04/09/07 | 43 | 02:56:46 | 06:54:17 | 03:57:31 | | - | P. Wiggins |
| 04/09/07 | 44 | 00:32:55 | 05:14:13 | 04:41:18 | | - | J. Pollock |

**Table 3:**
Observed magnitudes drops and times of eclipse ingress and egress for each recorded event. Predicted times are also given. The subscripts 1 and 2 respectively stand for Kalliope and Linus.

| Date | Type of event | Observed time of ingress (UTC) | Predicted time of ingress (UTC) | Observed time of egress (UTC) | Predicted time of egress (UTC) | Magnitude drop |
|---|---|---|---|---|---|---|
| 2007-03-08 | 1E2 | 1:15±1' | 2:32 | 3:59±1' | 5:06 | 0.050 |
| 2007-03-17 | 2E1 | 0:25±5' | 1:46 | 2:57±1' | 4:56 | 0.078 |
| 2007-03-27 | 2E1 | 23:05±5' | 21:00 | -- | 23:30 | ~0.050 ? |

**Table 4:**

The circular orbit solution adopted in the present work from Marchis et al. (2008a).

| Orbital parameters | |
|---|---|
| Period (days) | 3.5954±0.0011 |
| Semi-major axis (km) | 1099±11 |
| Pole solution in ECJ2000 (degrees) | λ= 197±2°, β= -3±2° |
| Inclination (degrees) | 99.4±0.5° |
| Ascending node (degrees) | 284.5± 2.0° |
| Time of pericenter (JD) | 2452185.489±0.006 |

**Table 5:** Astrometry of Linus with respect to the center of Kalliope at ingress and egress times as inferred from our events modeling.

| UTC Date | X (mas) | Y(mas) |
|---|---|---|
| 2007-03-08T1:15 | +220±1 | +68±4 |
| 2007-03-08T3:59 | +225±1 | -55±4 |
| 2007-03-17T0:25 | -233±1 | -63± 4 |
| 2007-03-17T3:15 | -234±1 | +55±4 |

**Table 6:**
Analysis of the radiometric measurements of IRAS using the NEATM (Harris, 1998). Given our knowledge of the shape of Kalliope we can calculate the effective diameter at the time of the IRAS sightings (Fig. 11 and text). Using the 3-D model average diameter of 166 km, we obtain an effective diameter at the time of the IRAS sightings in agreement with the NEATM model, validating our analysis.

| Observations | Pv, η | $D_{NEATM}$ in km | Pole-on effective diameter[2] with $D_{mean}$=166 km in km |
|---|---|---|---|
| IRAS1 1983/ 06/ 15 | 0.15, 0.74 | 175. | 181 |
| 179IRAS2 1983/06/15 | | 0.20, 0.43 | |
| IRAS3152[1] 1983/06/19 | 0.15, 0.72 | 174 | 181 |
| IRAS4 1983/06/19 | 0.14, 0.80 | 182 | 180 |

1. η unrealistic. Probably erroneous
2. Calculated on the basis of the 3D-shape model; contribution of the Kalliope I Linus satellite was removed

**Table 7**

List of the Japanese people with the geographic coordinates of the places from where they have observed the stellar occultation by Kalliope.

| #ID | Observer's name | East Longitude | North Latitude | Altitude |
|---|---|---|---|---|
| | | ° ' " | ° ' " | m |
| 1 | H. Sato | 140 29 25.3 | 37 44 35.9 | 90 |
| 2 | M. Kashiwagura | 140 08 40.2 | 38 22 47.5 | 182 |
| 3 | H. Tomioka | 140 41 09.0 | 36 38 33.0 | 33 |
| 4 | A. Yaeza | 140 36 11.0 | 36 31 28.0 | 230 |
| 5 | H. Okita et al. | 140 39 24.7 | 38 13 10.2 | 300 |
| 6 | M. Koishikawa et al. | 140 45 53.3 | 38 16 19.9 | 100 |
| 7 | M. Satou | 140 59 12.8 | 37 24 03.8 | 58 |
| 8 | S. Uchiyama | 140 32 11.5 | 36 20 36.3 | 6 |
| 9 | S. Suzuki | 139 32 04.6 | 35 22 44.2 | 30 |
| 10 | M. Sato | 139 28 33.7 | 35 40 54.5 | 60 |
| 11 | R. Aikawa | 139 26 50.4 | 35 56 56.5 | 21 |
| 12 | M. Yanagisawa | 139 32 37.0 | 35 39 27.7 | 80 |
| 13 | T. Tanaka | 139 34 50.0 | 35 17 53.0 | 9 |
| 14 | K. Kitazaki | 139 33 41.2 | 35 42 36.9 | 66 |
| 15 | E. Katayama | 139 33 22.2 | 35 41 52.9 | 62 |
| 16 | T. Ohkawa | 139 31 03.2 | 35 35 13.3 | 39 |
| 17 | T. Hayamizu | 130 18 00.7 | 31 50 19.2 | 20 |
| 18 | H. Takashima | 139 58 06.0 | 35 49 51.0 | 28 |
| 19 | H. Fukui | 138 15 14.0 | 34 51 05.0 | 25 |
| 20 | H. Suzuki | 137 42 48.4 | 34 45 53.8 | 50 |
| 21 | K. Kenmotsu | 133 44 52.6 | 34 41 40.3 | 53 |
| 22 | A. Asai | 136 31 24.3 | 35 10 14.2 | 187 |
| 23 | A. Hashimoto | 139 01 59.6 | 35 58 04.5 | 355 |
| 24 | A. Matsui | 138 15 31.4 | 36 18 31.9 | 557 |
| 25 | S. Uehara | 140 07 04.2 | 36 05 11.2 | 30 |

**Table 8:** Timings of the disappearance (D) and reappearance (R) of the star recorded by the observers. The column 'Obs. Method' displays the observing mode: P.E. = photo-electrical; Vis. = visual.

| #ID | Disappearing time | Reappearing time | Duration | Accuracy D | Accuracy R | Obs. Method | Observer's comments |
|---|---|---|---|---|---|---|---|
| | h m s | h m s | s | s | s | | |
| 1 | 19 49 18.38 | 19 49 44.43 | 26.05 | 0.15 | 0.07 | P.E. | gradual: $0^s$.30D, $0^s$.13R |
| 2 | 19 49 37.30 | 19 49 54.47 | 17.17 | 0.05 | 0.07 | P.E. | gradual: $0^s$.10D, $0^s$.13R |
| 3 | 19 48 57.50 | 19 49 24.37 | 26.87 | 0.06 | 0.07 | P.E. | gradual: $0^s$.07D, $0^s$.10R |
| 4 | 19 48 57.70 | 19 49 22.10 | 22.40 | 0.20 | 0.30 | Vis. | |
| 5 | 19 49 27.80 | 19 49 53.60 | 25.80 | 0.20 | 0.20 | Vis. | |
| 6 | N.A. | 19 49 53.78 | 0.00 | 0.01 | 0.50 | P.E. | D cloudy |
| 7 | 19 49 08.50 | 19 49 36.98 | 28.48 | 1.50 | 0.50 | Vis. | |
| 8 | 19 48 57.64 | 19 49 18.02 | 20.38 | 0.04 | 0.04 | P.E. | |
| 9 | 19 47 56.10 | 19 47 59.91 | 3.81 | 0.04 | 0.04 | P.E. | |
| 10 | 19 48 02.50 | 19 48 05.70 | 3.20 | 0.30 | 0.20 | Vis. | |
| 11 | 19 48 06.80 | 19 48 10.60 | 3.80 | 0.30 | 0.30 | Vis. | D gradual |
| 12 | 19 48 00.46 | 19 48 05.37 | 4.91 | 0.10 | 0.10 | P.E. | gradual: $0^s$.1D, $0^s$.1R |
| 13 | 19 47 54.90 | 19 47 59.00 | 4.10 | 0.500 | 0.50 | Vis. | |
| 14 | 19 48 01.31 | 19 48 06.34 | 5.03 | 0.04 | 0.04 | P.E. | gradual: $0^s$.06D, $0^s$.06R |
| 15 | 19 48 01.00 | 19 48 06.00 | 5.00 | 0.30 | 0.30 | P.E. | |

# Figures

**Figure 1**

Collected photometric lightcurves, not light-time corrected, between November 2006 and March 2007. They were used to derive a new shape model from an inversion method (Kaasalainen et al.,2001). 3D shape model is rendered in Fig.2. The new J2000 ecliptic pole is λ = 195±3° and β = +7±2°. Synthetic lightcurves reckoned from such a model have been overlapped (solid lines). Small discrepancies of 0.02mag at most arise from March 2007 due to a phase angle greater than ~20° (see section 2 for supplementary explanations).

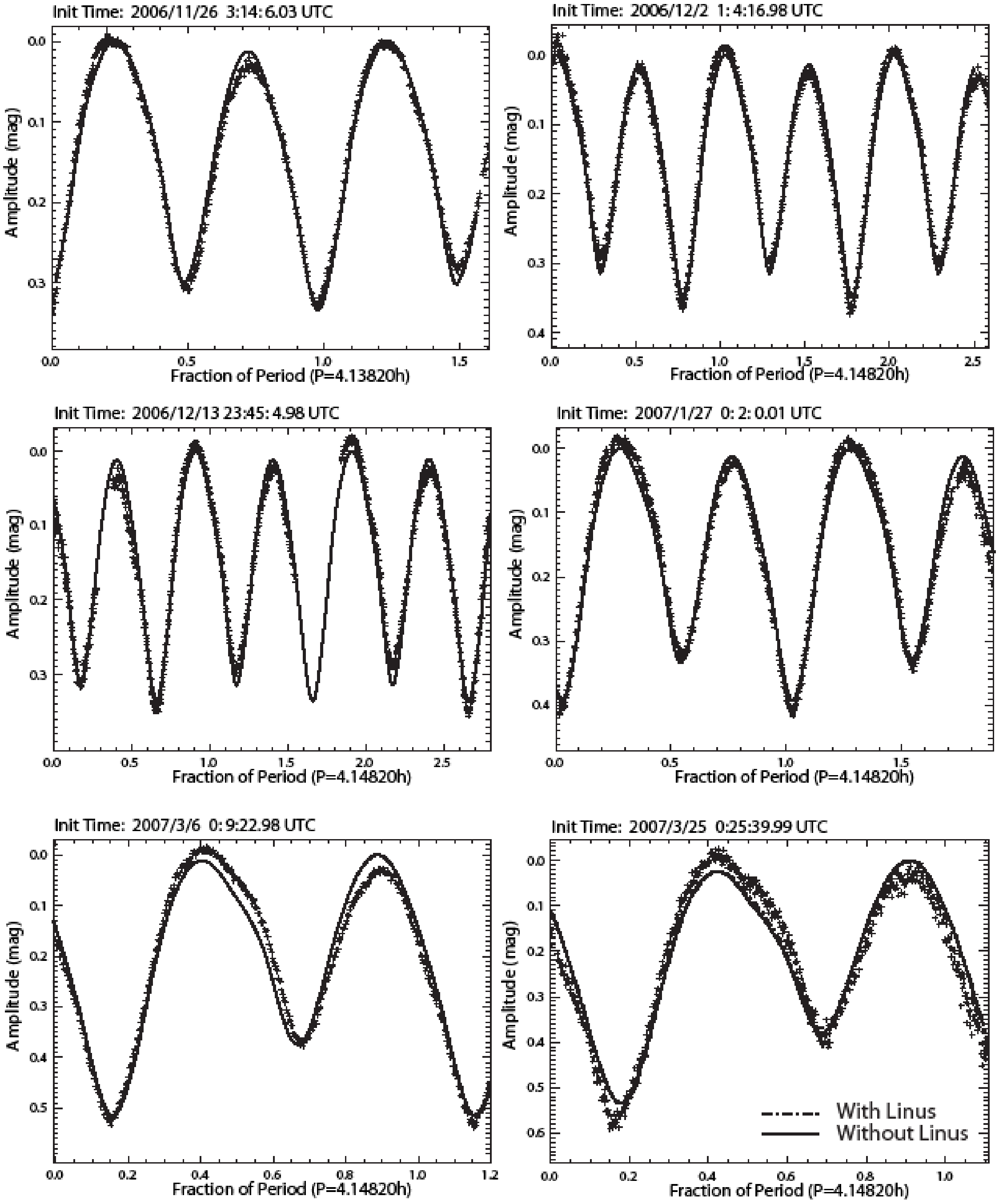

**Figure 2**
Side (*top*) and tip (*bottom*) views of the three-dimensional shape model of 22 Kalliope. As Kalliope was seen in a nearly edge-on aspect, hitherto never observed in such manner, the new model is consequently improved over the preceding one (Kaasalainen et al., 2002). Kalliope is slightly more elongated, accounting for a peak-to-peak amplitude ranging from 0.35 to 0.6mag between November 2006 and April 2007. The $J_2$ term of the corresponding gravitational potential is of 0.19 assuming a uniform mass distribution.

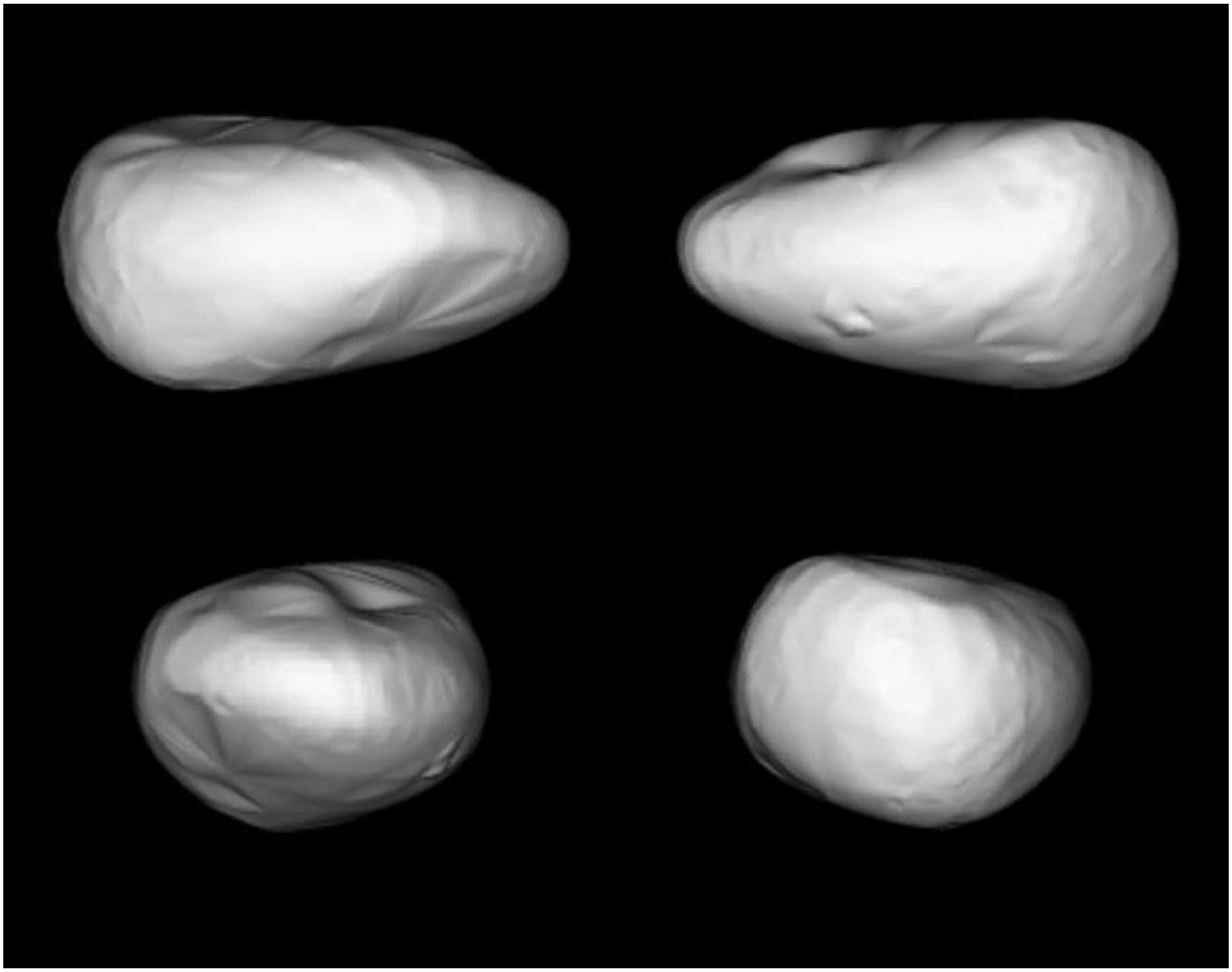

## Figure 3

Collected lightcurves not corrected for light-time. See Table 2 for details on each observation, identified by its ID number.

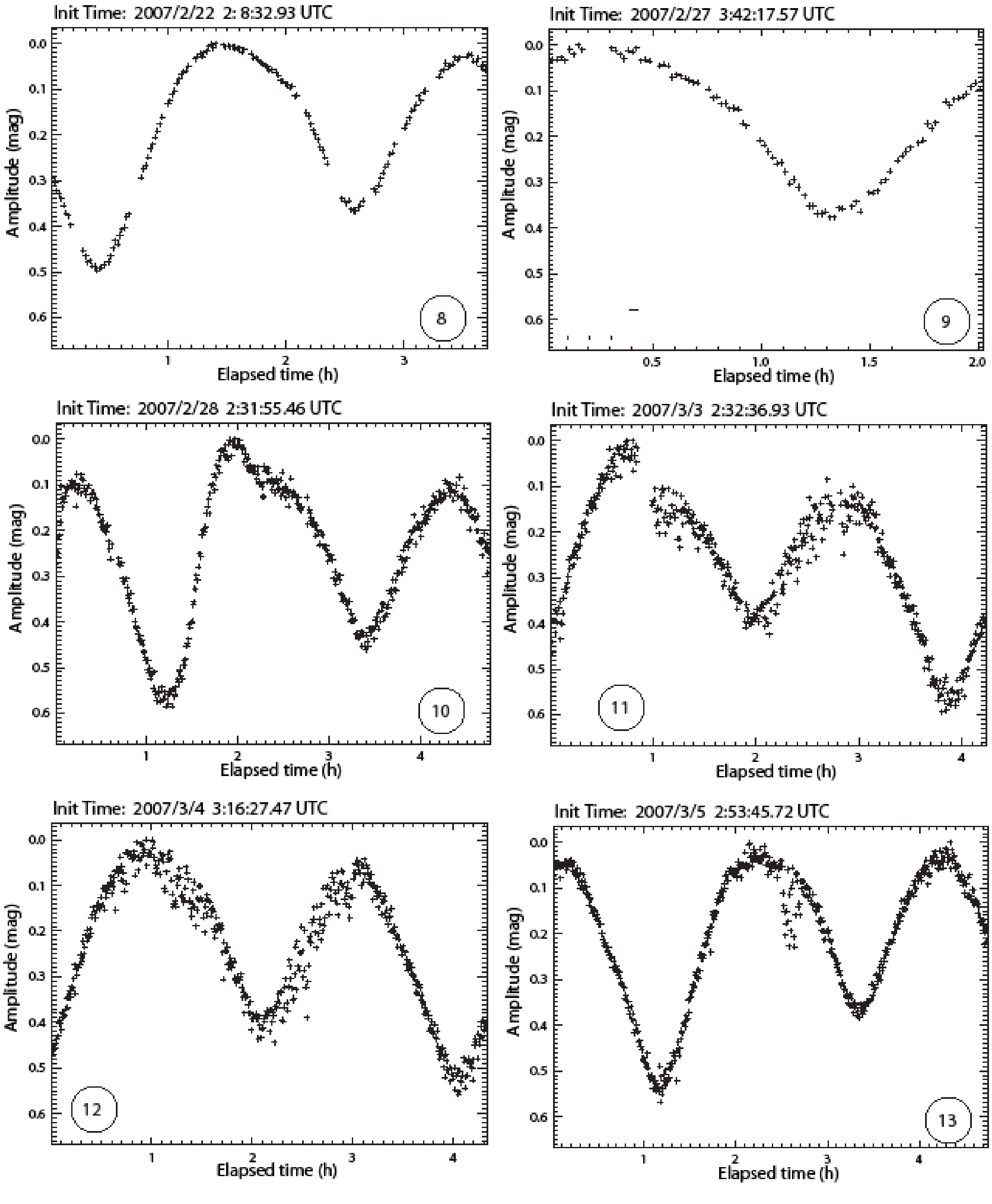

**Figure 3- continued**

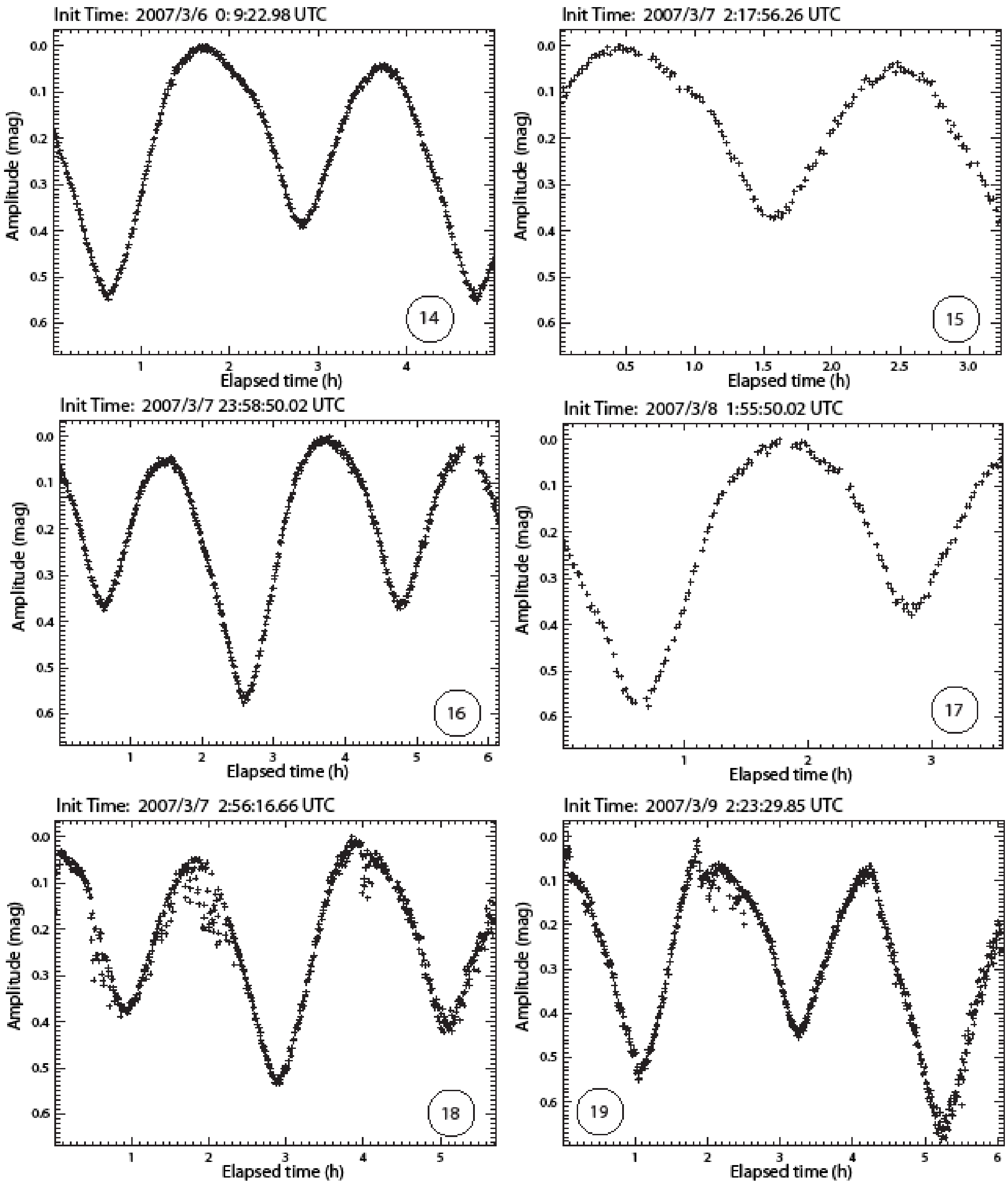

**Figure 3 - continued**

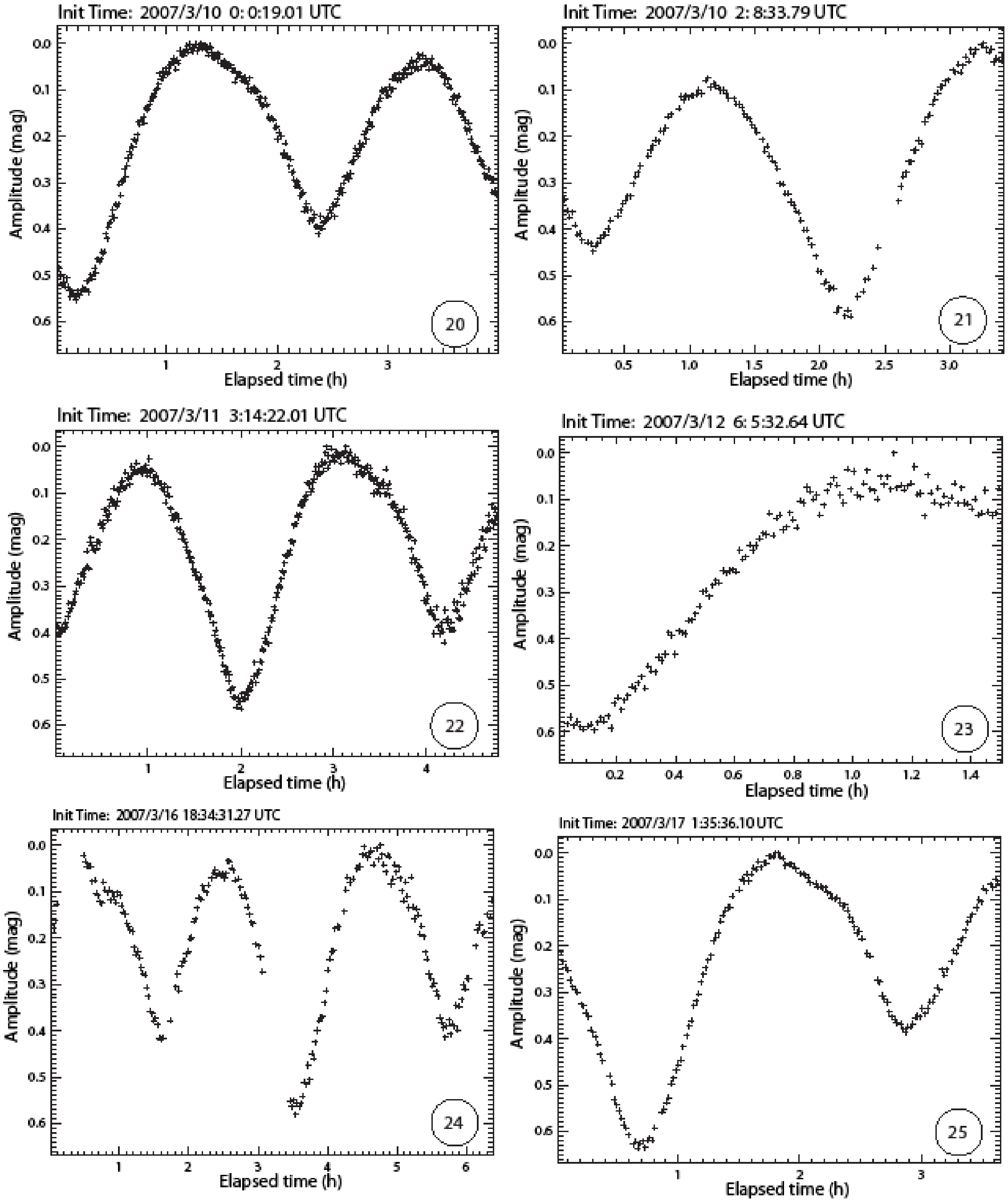

**Figure 3 - continued**

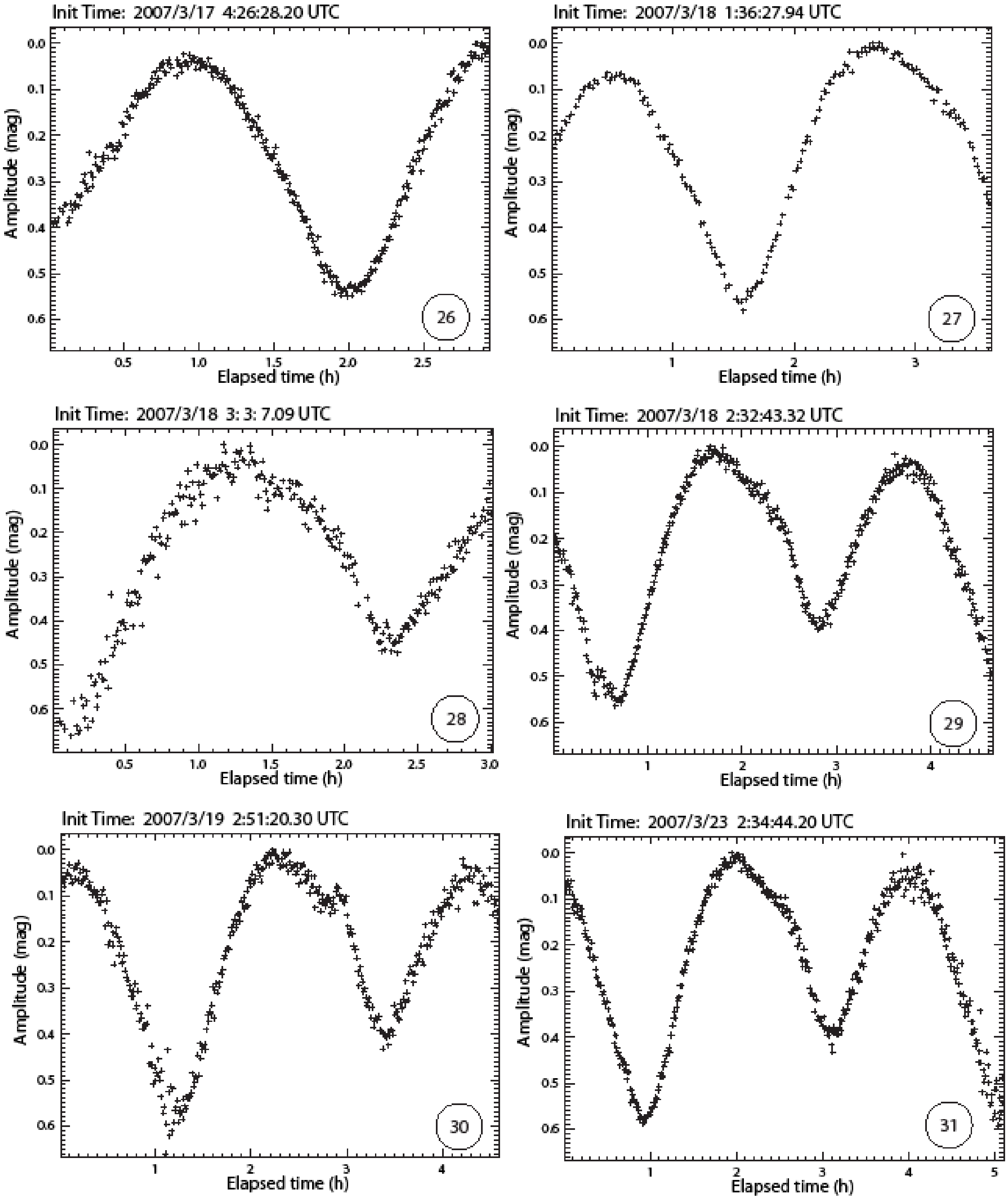

**Figure 3 - continued**

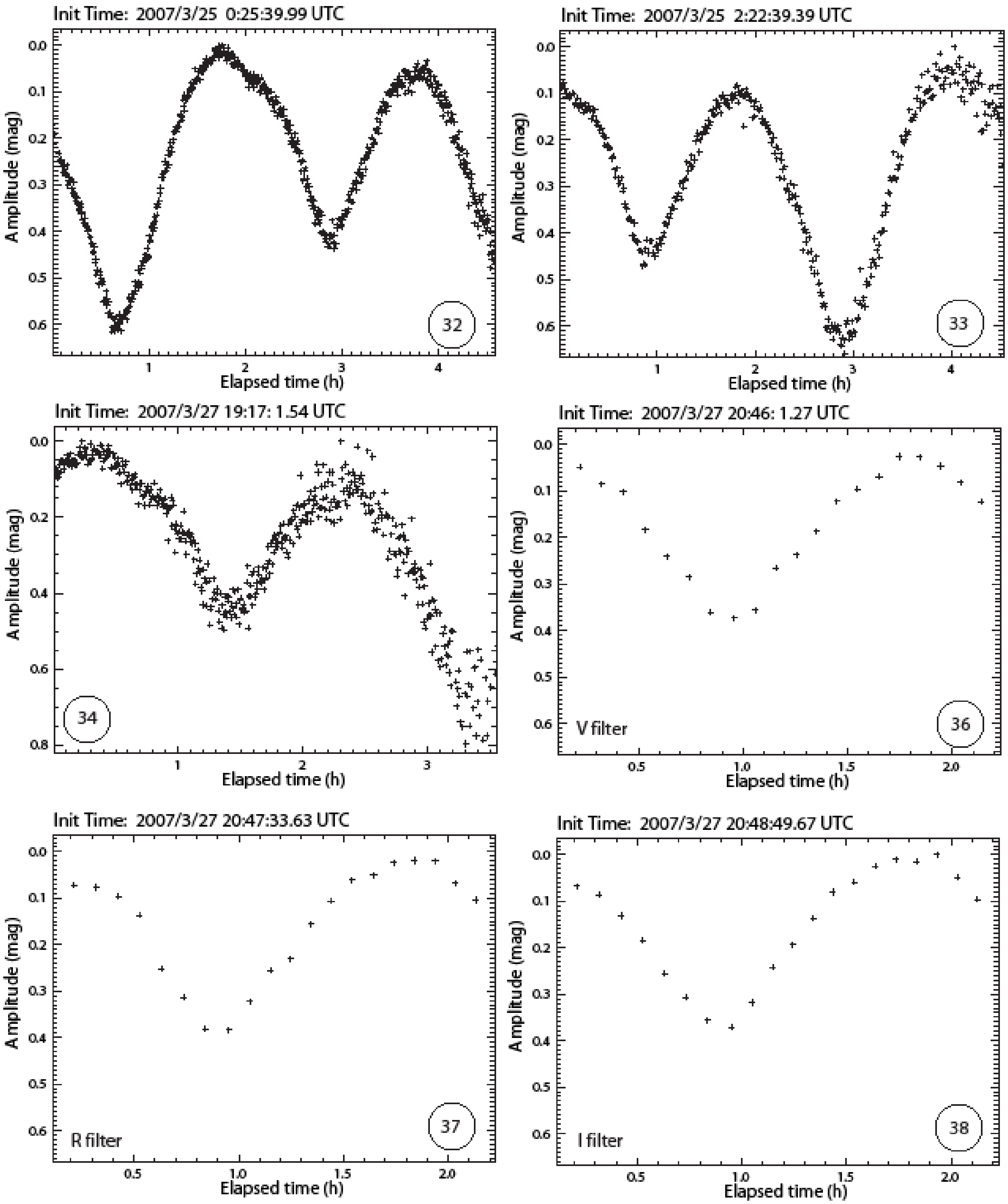

**Figure 3 - continued**

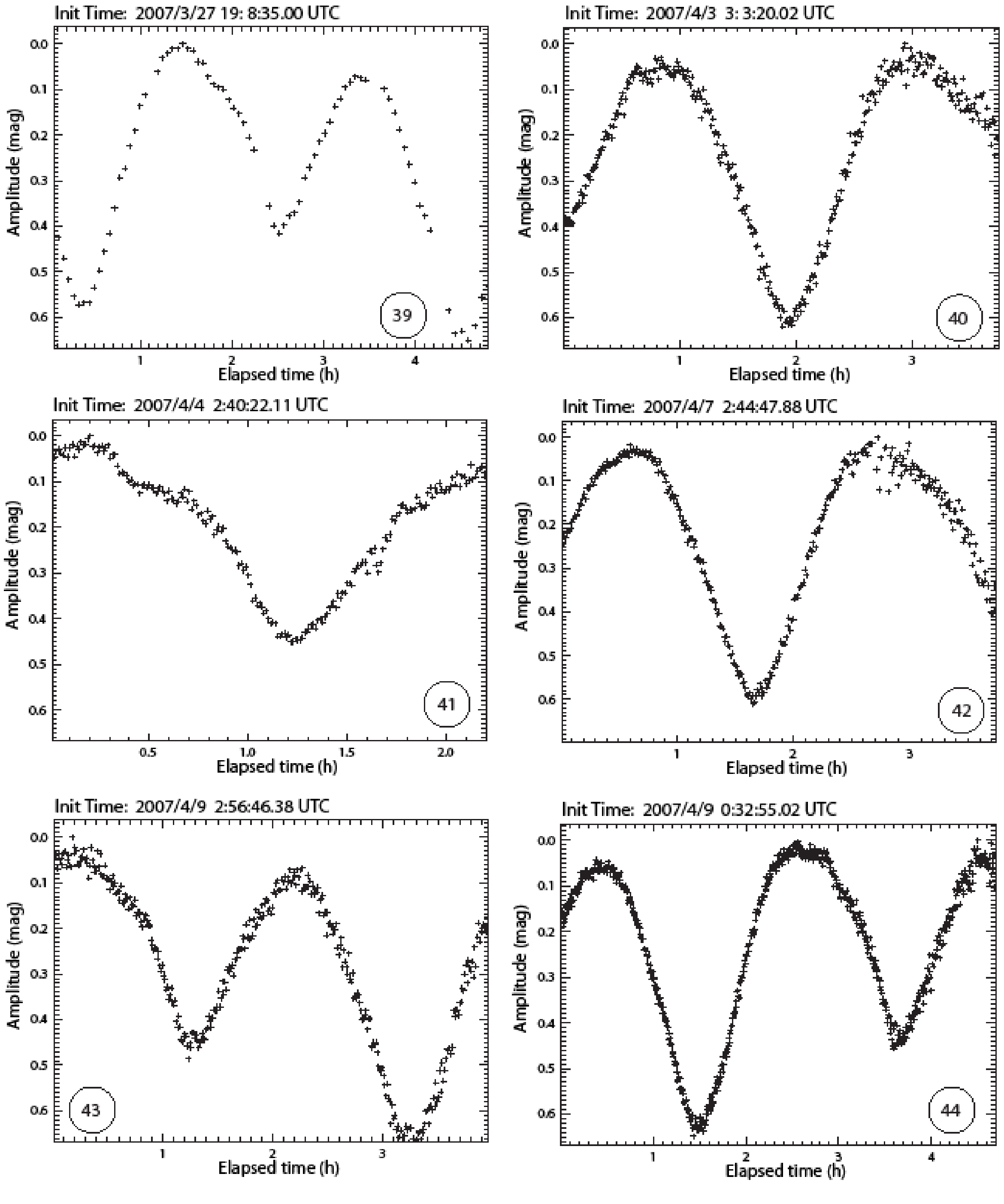

## Figure 4

Positive detections of a total eclipse of Linus by Kalliope on 2007 March, 8. To illustrate the detection of an event, a reference lightcurve obtained near in time (#14) is overlapped in the panels on the left (solid curve). The difference between the two lightcurves is plotted on the right side (see text for details). The identification number of the involved runs refers to the ones given in table 2. The detection is performed from two different sites (#16 and #17) which ascertains its reliablity.

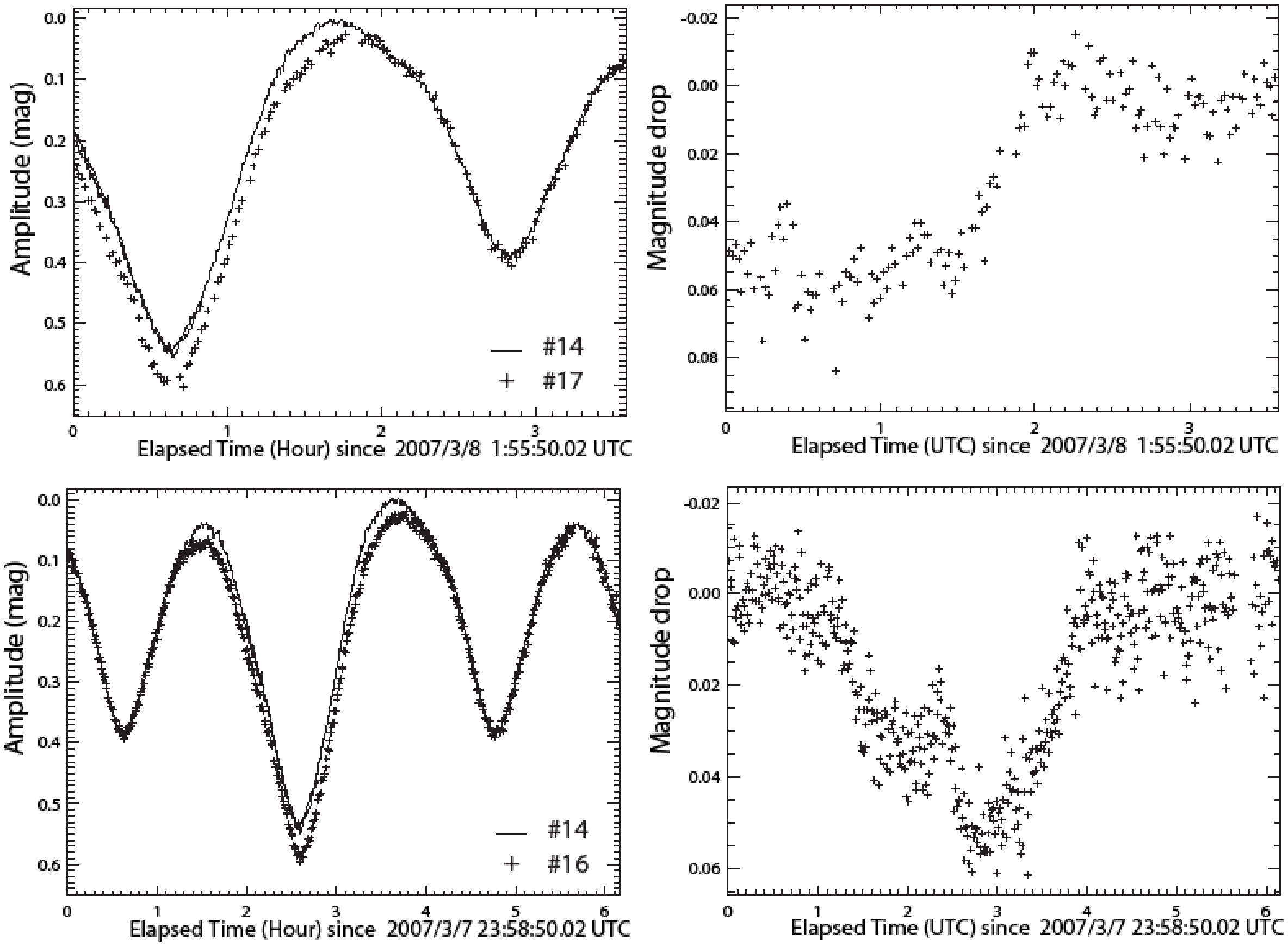

## Figure 5

Positive detections of a total eclipse of Kalliope by Linus on 2007 March, 17. The eclipse ingress is detected on the run #24 while the run #25 caught the end of the event. The two lightcurves were merged to get a full event.

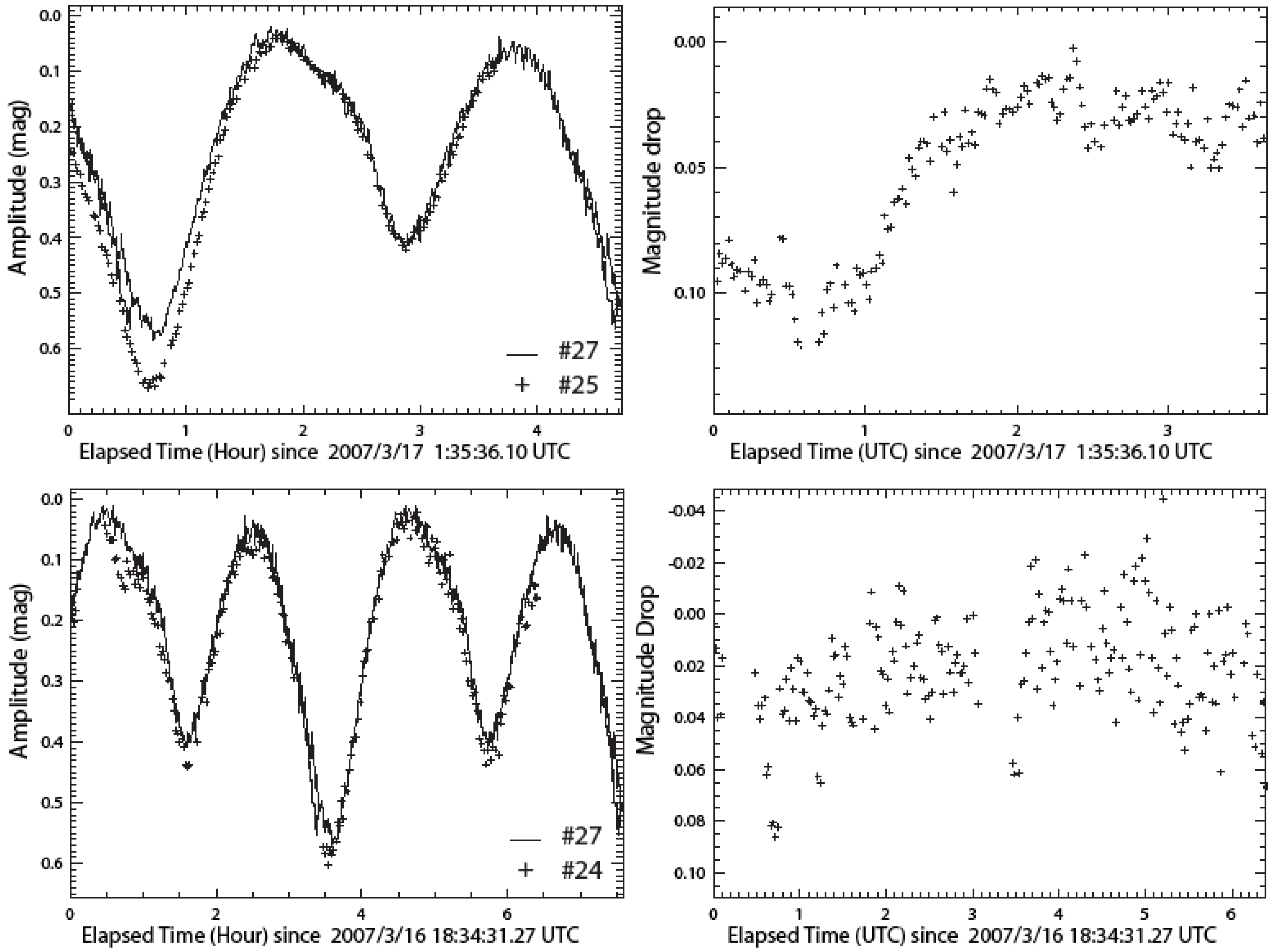

## Figure 6

Positive detections of a total eclipse of Kalliope by Linus on 2007 March, 27. Only the eclipse ingress was observed.

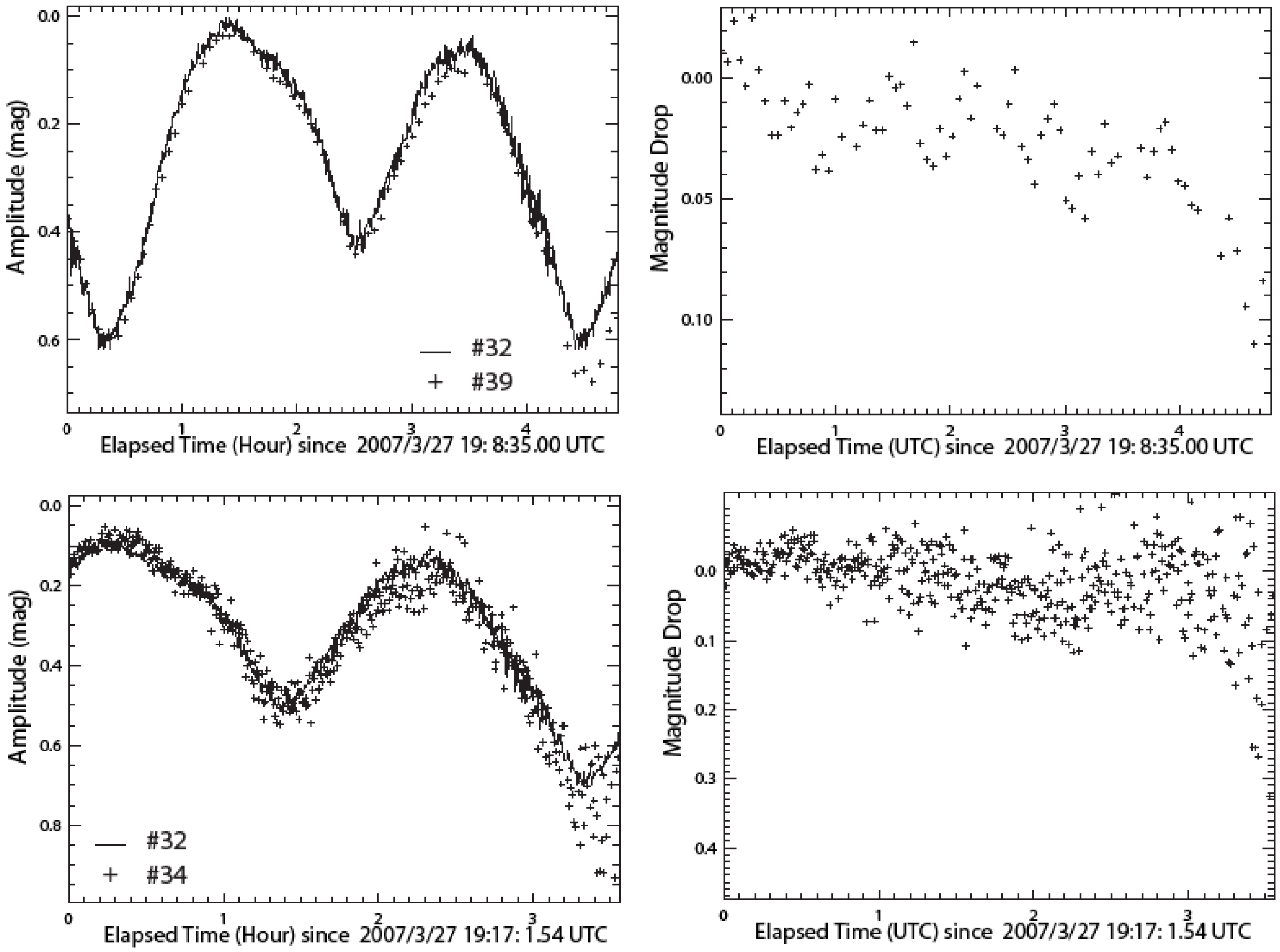

**Figure 7**
Viewing geometry of the eclipse of Kalliope by Linus on March 17 2007. Linus with its path around Kalliope are visible on the right. The event is shown at three times of which the time of midevent (1:48 UTC) and the time of maximum flux drop (2:24 UTC). The cylindrical umbra of Linus may cover a variable fraction of Kalliope depending on its aspect. North is up and east is to the left.

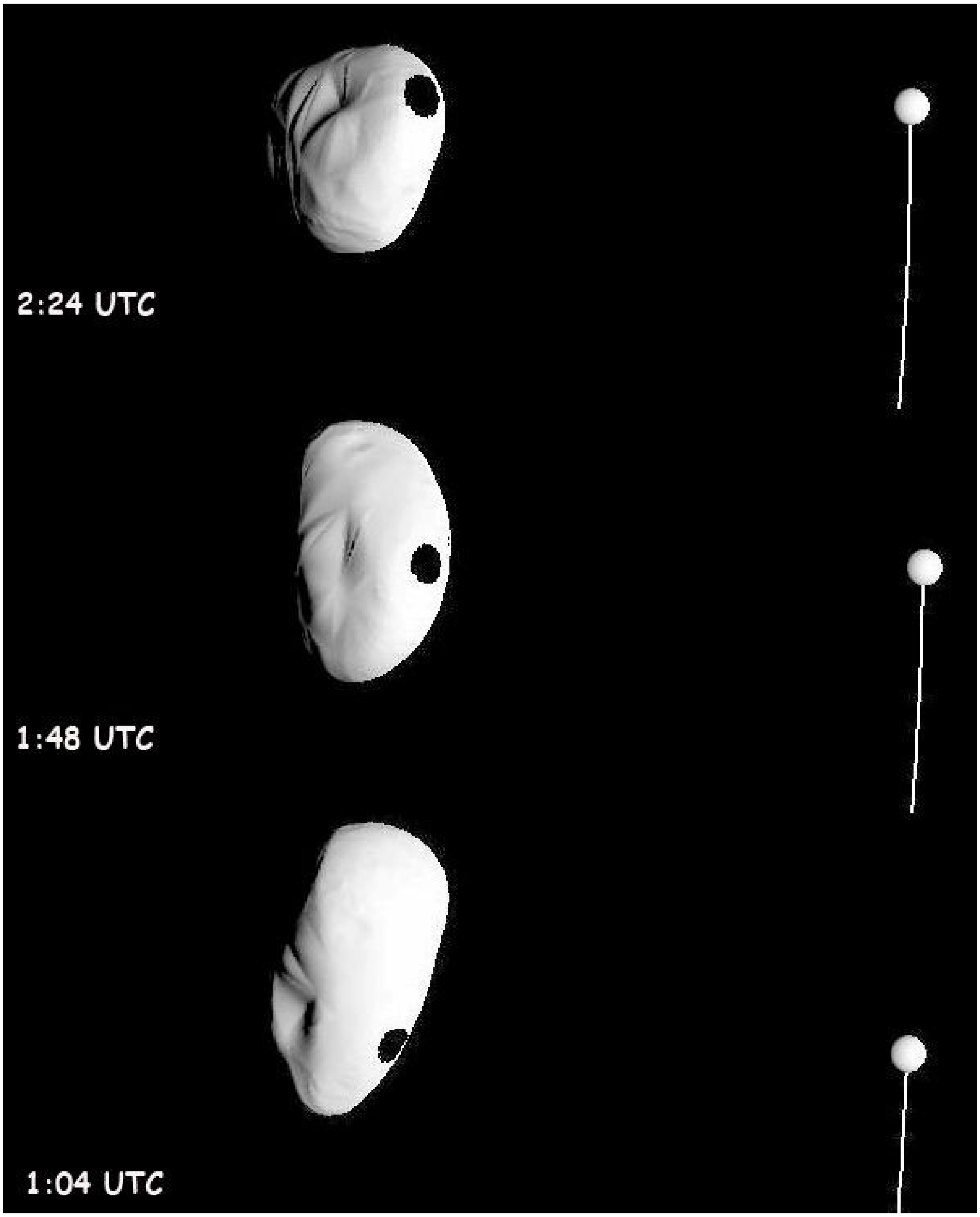

## Figure 8

Effect of the parameters on the drop curve. These simulations are carried out for the event on March 8 2007. Each parameter is varied while keeping others as constant. The parameters are the effective radius of Kalliope (A), the radius of Linus (B), the ecliptic coordinates of the orbit pole of Linus (C, D), the limb-darkening parameter (E) and the 3D shape model of Kalliope (F).

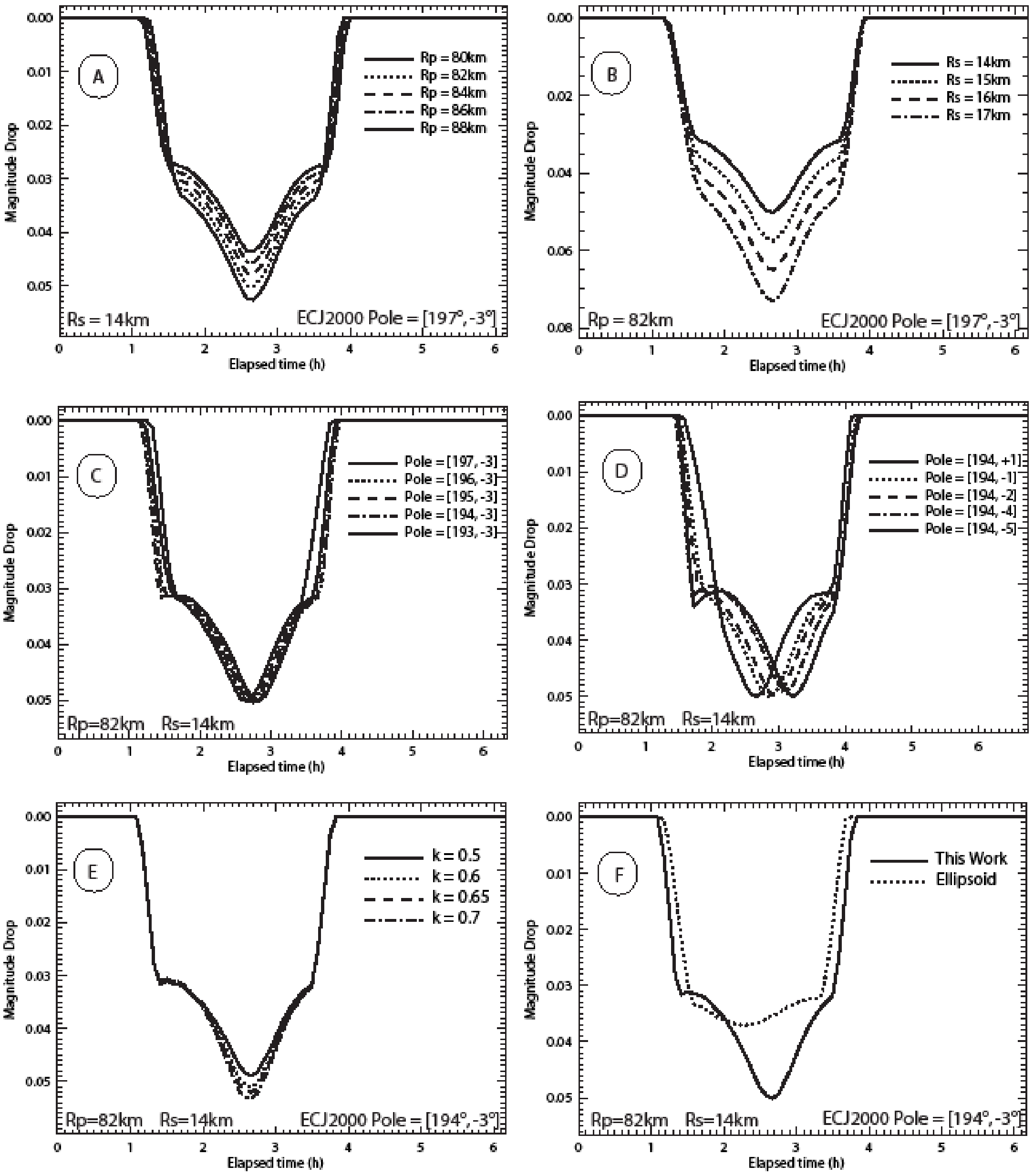

## Figure 9

Evolution of the duration of the total eclipse of Linus by Kalliope on March 8 as a function of the effective radius of Kalliope. The observed duration is 2.75±0.03h. The best fitted effective radius is derived to 83.1±1.4km. The IRAS radius of Kalliope (90.5km) is clearly inadequate to reproduce the required duration of the event.

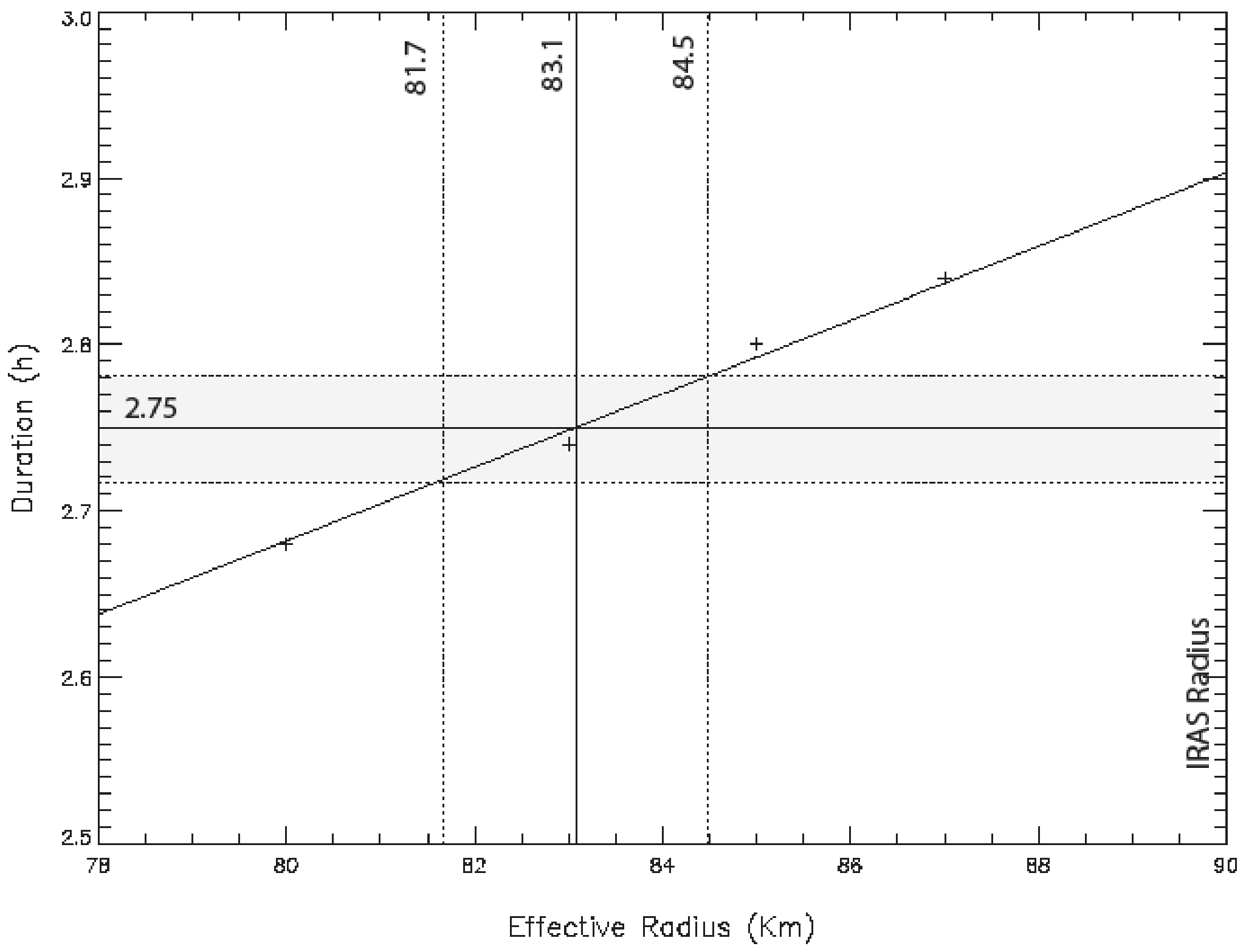

# Figure 10

Synthetic curves of magnitude drop corresponding to the best-fit solution are superimposed to the positive detections shown on the figures 4, 5 and 6.

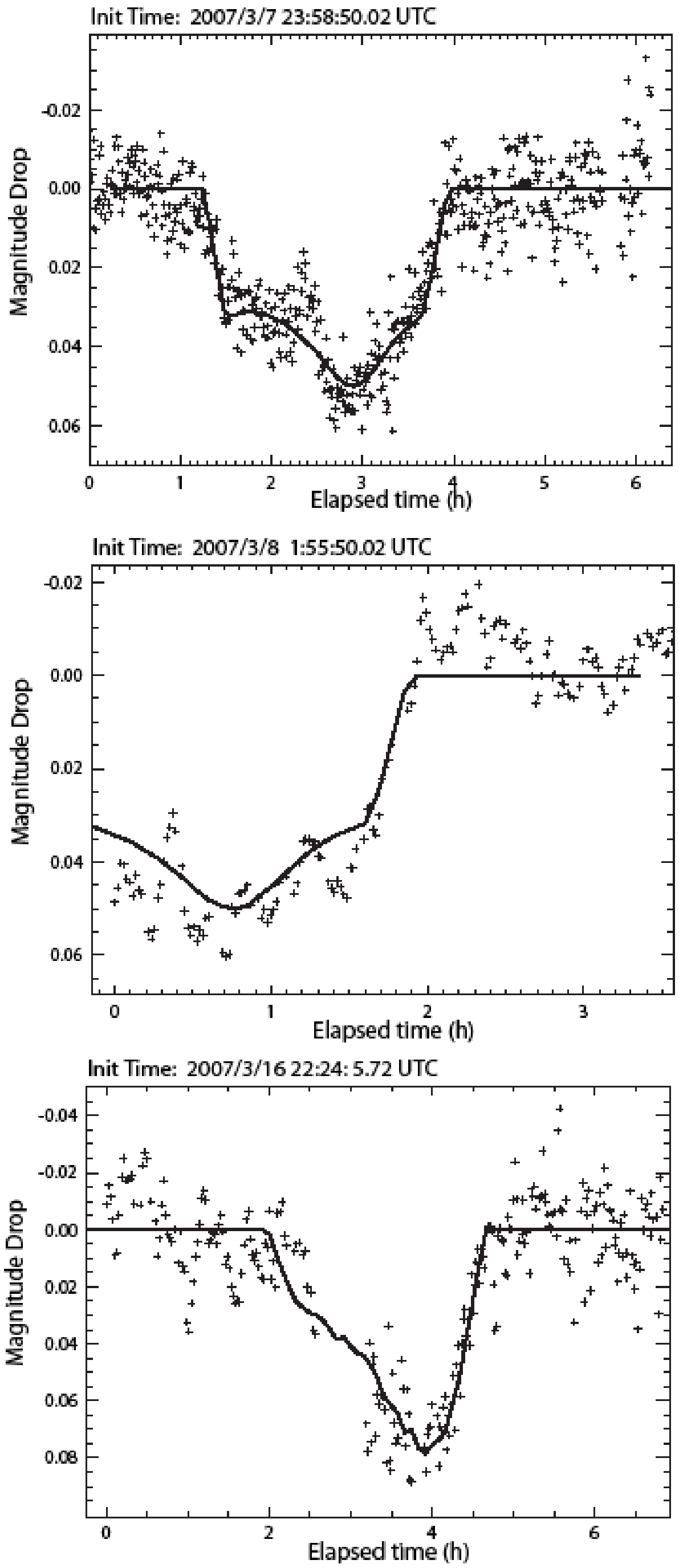

## Figure 11

Apparent projected shape of Kalliope's primary during the IRAS sightings (June 15,19 1983) obtained using our pole solution and 3D-shape model. The asteroid was observed pole-on and exhibited its maximum apparent shape. The diameter (180 km see Table 6) derived from the IRAS sightings is therefore significantly larger than the true mean diameter.

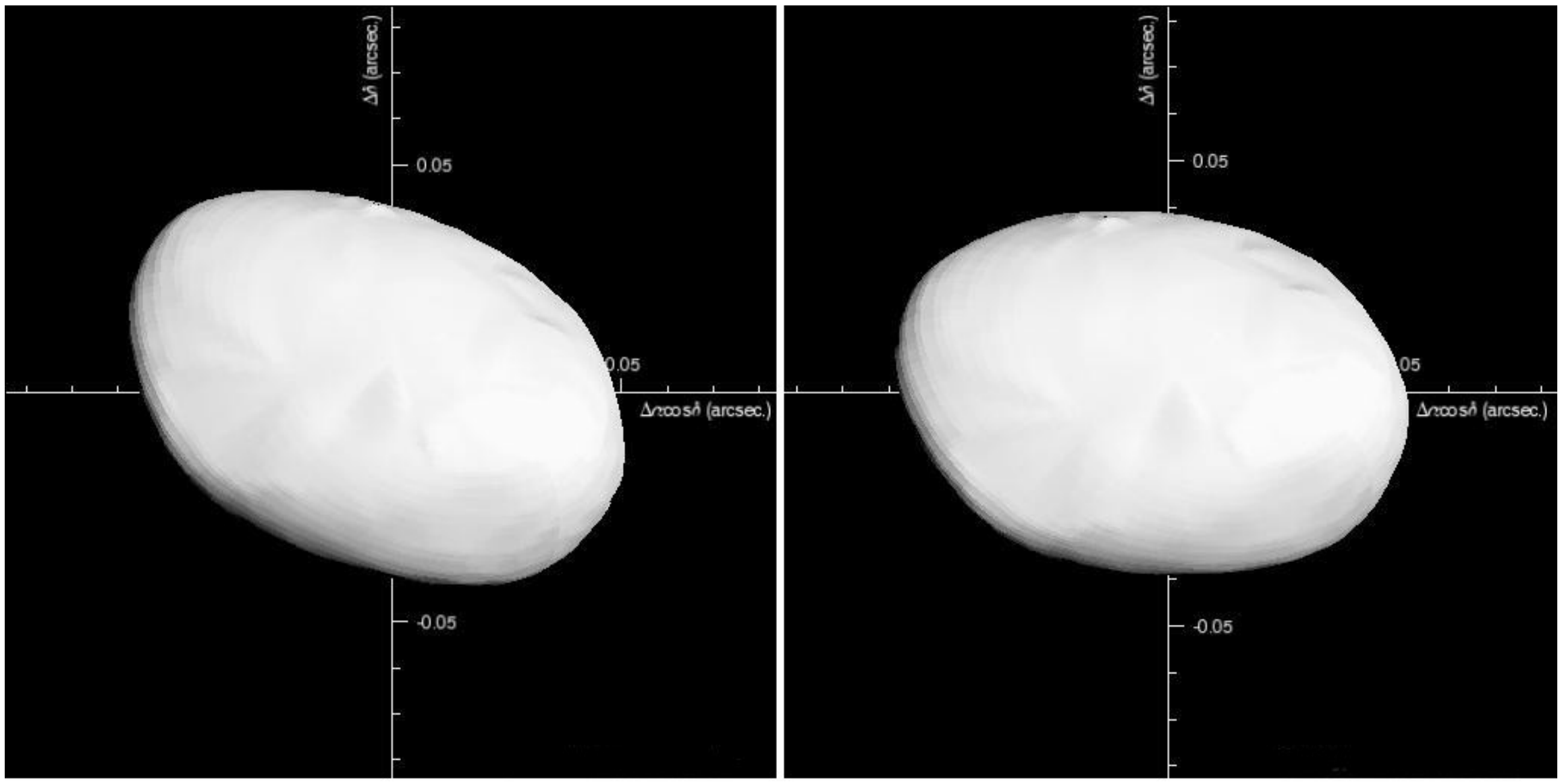

## Figure 12

Stellar occultation by Kalliope and its satellite Linus on November 7, 2006. Solid lines are the recorded chords for Kalliope and Linus during the stellar occultation. The dashed lines show the negative observations (i.e. no disappearance of the star); the cross materializes the predicted position of Linus. The departure from the observed position, derived after fitting an ellipsoidal outline (Fig.13), is of 10 ±10km (8±8 mas) along the x axis and 57 ±10 km (44±8 mas) along the y axis.

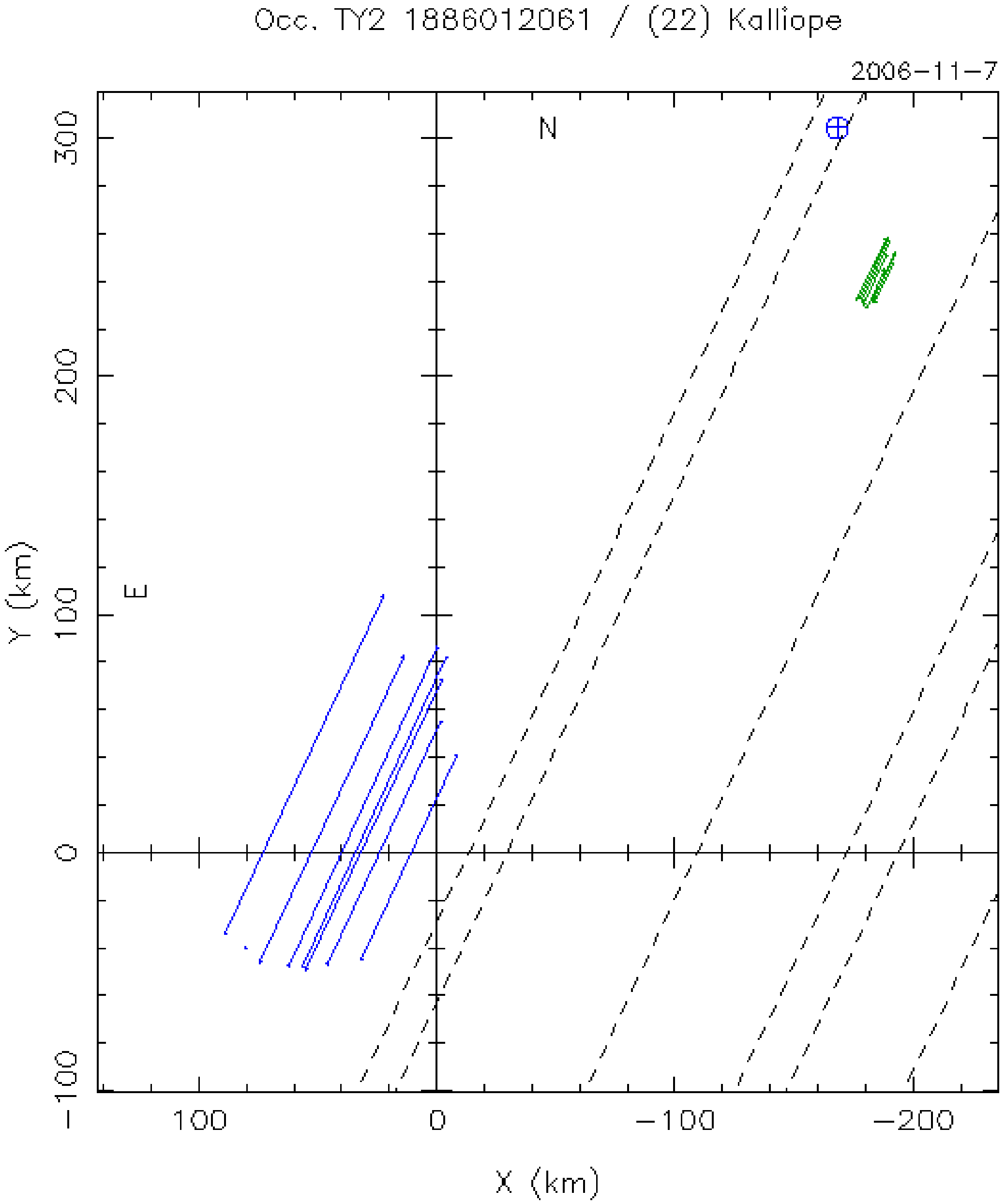

**Figure 13**

Linus profile (solid line with an equivalent diameter of 30±6 km) fitted on the observed chords. The dashed curves show the uncertainty of the fitted profile. The two dashed lines on the edges show negative detections. The semimajor axes are 16±5 km and 14±1 km with a position angle of 40.2°.

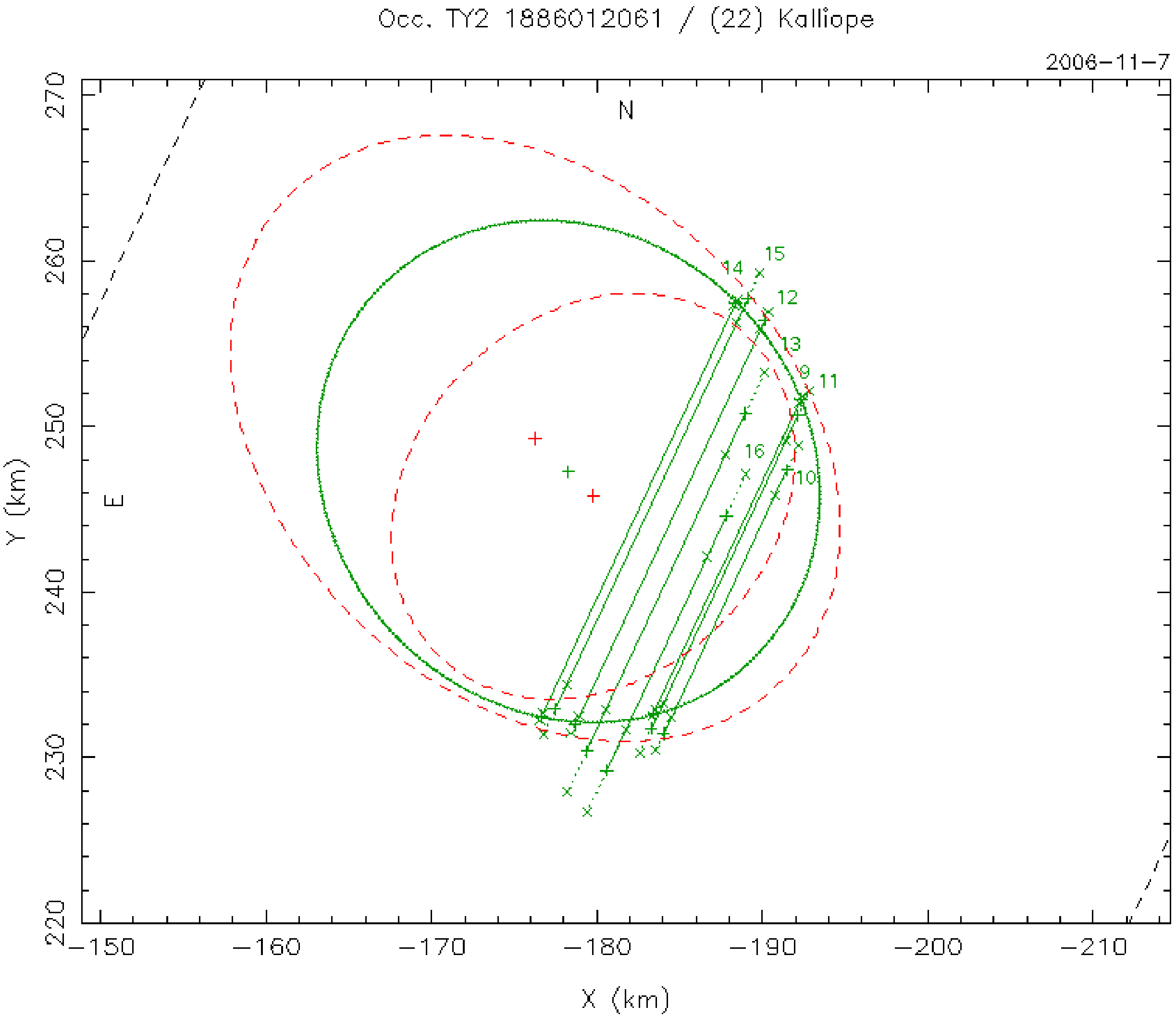

## Figure 14

Limb profiles of Kalliope for two equivalent diameters, 166km (this work, innermost outline) and 180km (IRAS value, outermost outline) adjusted to the chord set. North is up and east is to the left. The adjustment is carried out with the emersion points which are usually the most accurately determined (lower points). The shape and size solution of this present work give the best overall agreement with the observed chords. However, serious discrepancies are apparent with two points corresponding to the immersion timings of chords 5 and 7 which deviate from the projected limb profile by about 20km (or 1.1sec of time), an amount which may be explained by a significantly non-convex topography on Kalliope instead of a non-reasonable timing error.

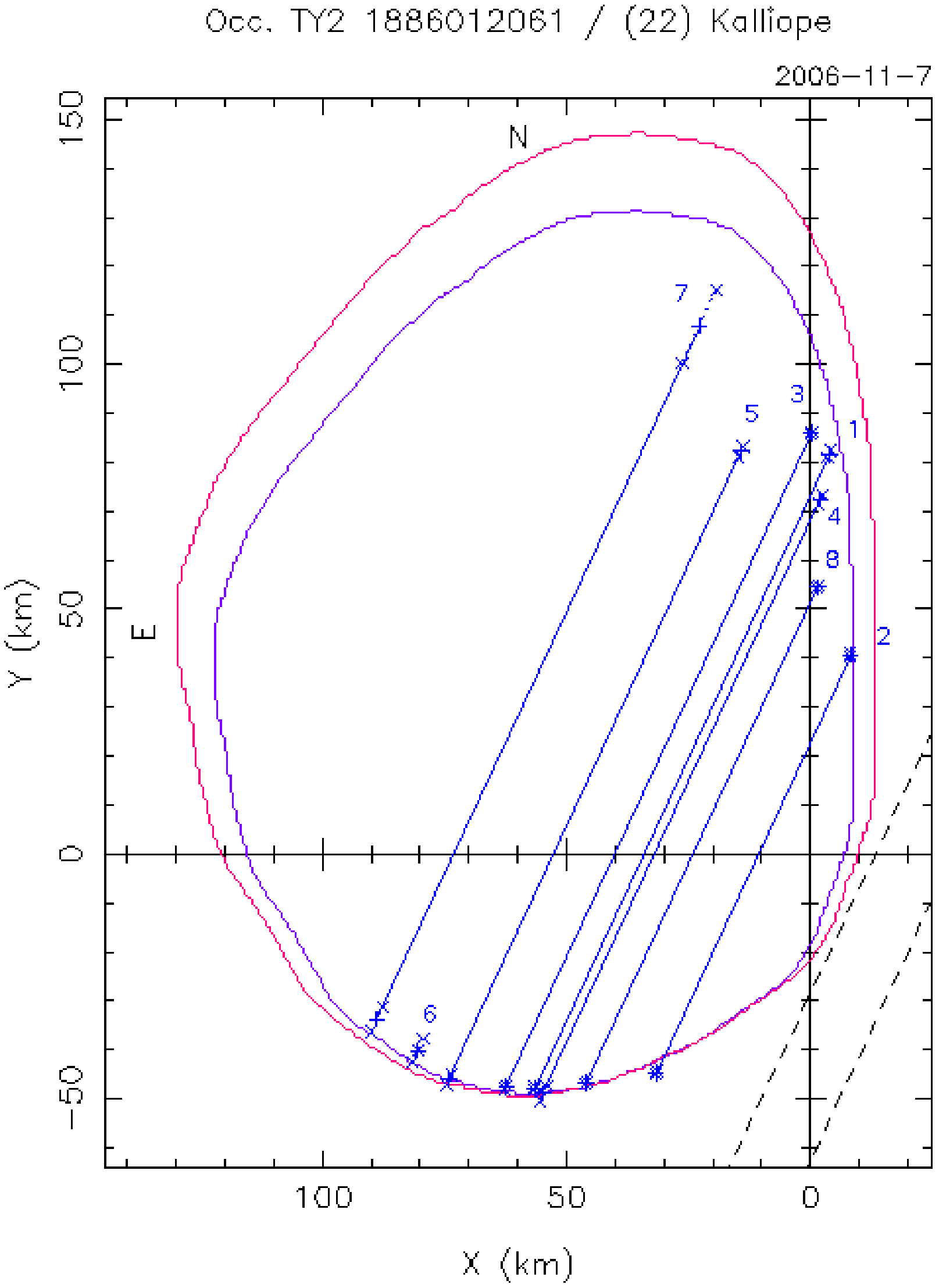

## Figure 15

Diagram of the relative separation a/Rp against the mass ratio q after Weidenschilling et al. (1989). Curves of a same evolution time are plotted. They assume initial a/R=1, $\mu Q \approx 3x10^{13}$dynes/cm$^2$, $\rho$=3.4g/cm$^3$, Rp=83.1km and q =0.0047 which are the physical characteristics of Kalliope derived in the present work. Binaries to the left of the curve labelled synchronous stability cannot maintain spin-orbit synchronism. The location of Kalliope in this diagram is indicative of a very primitive body with a probable origin in the early Solar System.

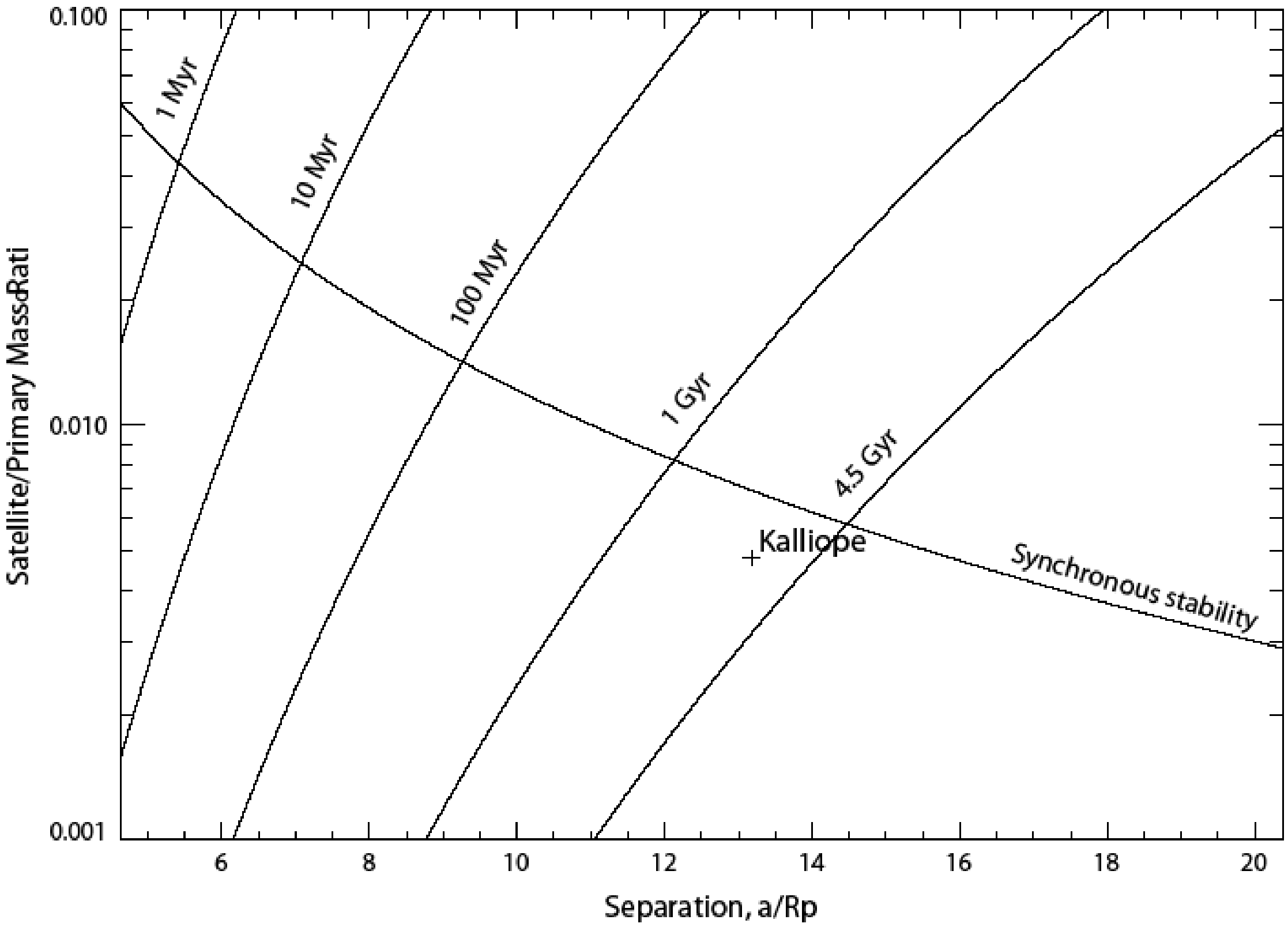